\documentclass[aps,twocolumn,showpacs,showkeys,preprintnumbers,amsmath,amssymb,amsfonts,nofootinbib]{revtex4}

\pdfoutput=1

\usepackage{subfigure}
\usepackage{graphicx}
\usepackage{color}
\usepackage{slashed}
\usepackage{tabularx}

\usepackage{extarrows}
\usepackage{dsfont}

\usepackage{hyperref}
\usepackage{url}
\usepackage{doi}
\newcommand{\Tr}{\mathrm{Tr}}
\newcommand{\tr}{\mathrm{tr}}

\newcommand{\I}{\mathrm{i}}
\newcommand{\Nf}{N_{\text{f}}}
\newcommand{\Nc}{N_{\text{c}}}

\newcommand{\dt}{\!\cdot\!}

\newcolumntype{L}[1]{>{\raggedright\arraybackslash}p{#1}} 
\newcolumntype{C}[1]{>{\centering\arraybackslash}p{#1}} 
\newcolumntype{R}[1]{>{\raggedleft\arraybackslash}p{#1}} 
\graphicspath{{.}{figures/}{mma/}}

\definecolor{blue}{rgb}{0,0,1}

\definecolor{green}{rgb}{0,1,0}

\definecolor{red}{rgb}{1,0,0}

\def\Eq#1{Eq.~(\ref{#1})}
\def\Sec#1{Sec.~\ref{#1}}
\def\Fig#1{Fig.~\ref{#1}}
\def\App#1{App.~\ref{#1}}
\def\Tab#1{Tab.~\ref{#1}}

\begin{document}

\title{Fluctuating vector mesons in analytically continued FRG flow equations}

\newcommand{\JLU}{Institut f\"ur Theoretische Physik, Justus-Liebig-Universit\"at, 35392 Giessen, Germany}

\author{Christopher Jung}
\author{Lorenz von Smekal}

\affiliation\JLU

\begin{abstract}
  In this work we study contributions due to vector and axial-vector meson fluctuations to their in-medium spectral functions in an effective low-energy theory inspired by the gauged linear sigma model. In particular, we show how to describe these fluctuations in the effective theory by massive (axial-)vector fields in agreement with the known structure of analogous single-particle or resonance contributions to the corresponding conserved currents. The vector and axial-vector meson spectral functions are then computed by numerically solving the analytically continued functional renormalization group (aFRG) flow equations for their retarded two-point functions at finite temperature and density in the effective theory. We identify the new contributions that arise due to the (axial-)vector meson fluctuations, and assess their influence on possible signatures of a QCD critical endpoint and the restoration of chiral symmetry in thermal dilepton spectra.
\end{abstract}

\pacs{05.10.Cc, 12.38.Aw, 11.10.Wx, 11.30.Rd, 14.40.Be}
\keywords{keywords}
\keywords{vector mesons, spectral functions, functional renormalization group, chiral symmetry restoration}


\maketitle

\section{Introduction}\label{sec:intro}


Understanding the phases of strong-interaction matter and the transitions between them is still one of the main goals in theoretical and experimental heavy-ion physics \cite{Braun-Munzinger2009,Fukushima:2010bq,Shuryak:2014zxa,Andronic:2017pug}. In order to relate the experimental observations from heavy-ion collision experiments at various beam energies to the phase diagram of Quantum Chromodynamics (QCD) photons and dileptons are of particular importance to obtain information, e.g., on the temperature and the lifetime of the fireball \cite{vanHees:2015bna,Galatyuk:2015pkq,Rapp:2016xzw}. They arise from various processes and are emitted in all stages of the collisions. Because they escape the fireball almost unaffected, these electromagnetic probes carry the spectral information of the matter in the interior of the collision zone during the evolution of the fireball. Therefore, the spectral properties of strong-interaction matter and the features of the QCD phase diagram, such as the in-medium modifications of hadrons reflecting chiral symmetry restoration and the transition to the quark-gluon plasma, are encoded in the measured photon and dilepton spectra \cite{Rapp:1999ej}.

The thermal rates are determined by the in-medium electromagnetic spectral function and can therefore be used to probe the QCD phase diagram \cite{Rapp:2013ema}. Especially in the resonance region, for invariant masses of dilepton pairs below 1~GeV, the electromagnetic spectral function is dominated by vector mesons, in particular the $\rho$ meson which gave rise to the famous vector meson dominance model (VMD) \cite{Schildknecht:2005xr}. Chiral symmetry on the other hand requires one to study the $\rho$ alongside with its chiral partner, the $a_1$ meson.
From the in-medium modifications to the spectral functions of vector and axial-vector mesons one then hopes to deduce the nature of chiral symmetry restoration and to find evidence of an associated critical endpoint in the phase diagram of QCD \cite{Rapp:2013nxa,Rapp:2009yu,Hohler:2015iba,Galatyuk:2015pkq}.

In this paper we compute the in-medium spectral functions of vector and axial-vector mesons in an extended linear sigma model with quarks as a chiral low-energy effective theory for two-flavor QCD. As in a previous study \cite{Jung:2016yxl} we employ the functional renormalization group (FRG) as the non-perturbative computational framework to include fluctuations beyon mean-field approximations. The FRG has proven to be a powerful tool in diverse areas of physics, see \cite{Gies:2006wv,Berges:2000ew,Pawlowski:2005xe,Delamotte:2007pf,Bagnuls:2000ae} for reviews. One particular bonus that it shares with other functional methods here is its applicabilty, at least in principle,  to the entire  QCD phase diagram including the region of finite baryon density \cite{Pawlowski:2010ht,Strodthoff:2016dip} where lattice QCD suffers from the fermion sign problem. Mainly for technical reasons, applications to full QCD \cite{Braun:2014ata,Cyrol:2017ewj} are nowadays often been used to motivate effective low-energy models with the potential to constrain the input parameters at their upper limit of applicability, their ultraviolet (UV) scale. From this input, these can then be used to describe chiral symmetry restoration \cite{Schaefer:2004en,Tripolt:2017zgc} or model the deconfinement transition \cite{Schaefer:2007pw,Fukushima:2003fw} with two or three quark flavors \cite{Schaefer:2013isa, Rennecke:2016tkm,Fu:2019hdw}, and to describe vector mesons \cite{Rennecke:2015eba,Eser:2015pka}.

For real-time quantities such as spectral functions or transport coefficients one faces the additional problem, common to all Euclidean approaches to quantum field theory, of the analytic continuation from Euclidean back to Minkowski spacetime. Unlike lattice QCD, where numerical reconstruction methods are usually required, the FRG flow equations for correlation functions can be analytically continued before they are being solved \cite{Floerchinger2012,Kamikado2014,Pawlowski:2015mia}. For a comparison of spectral functions from such analytically continued aFRG flows with those from various reconstruction methods, see \cite{Tripolt:2018xeo}. The power of the aFRG flow equations has by now been demonstrated in various model studies where mesonic and fermionic spectral functions have been computed \cite{Tripolt2014,Tripolt2014a,Pawlowski:2017gxj,Tripolt:2018jre,Tripolt:2018qvi}, recently also self-consistently \cite{Strodthoff:2016pxx}. In order to assess truncation effects, one can furthermore compare with classical-statistical lattice simulations of spectral functions \cite{Berges:2009jz,Schlichting:2019tbr} which are able to capture exactly their universal critical behavior in the vicinty of continuous phase transitions.

Extending the previous study of aFRG flows for vector and axial-vector meson spectral functions \cite{Jung:2016yxl}, in this paper we include the contributions from the (axial-)vector mesons themselves to these aFRG flows.  The structure of the propagators for the massive (axial-)vector fields of the effective theory is thereby fixed via current-field identities, i.e.~requiring it to agree with that of analogous single-particle or resonance contributions to the corresponding conserved currents. In paricular, this implies that the massless single-particle contributions of the Stueckelberg formulation must be avoided. To implement such a structure into the FRG framework, we have to go beyond the leading order in the derivative expansion. Additional longitudinal modes are included which switch themselves off in the infrared, ensuring the transversality of the massive (axial-)vector correlations in this limit. To demonstrate the feasibility of this formulation, we compute the $\rho$ and $a_1$ spectral functions from the resulting aFRG flow equations at finite temperature and density, across the critical endpoint (CEP) in the phase diagram of the model, in parallel with the previous study in \cite{Jung:2016yxl} but now with the fluctuations due to the massive vector and axial-vector mesons included.  We identify the new contributions that arise from these fluctuations, and assess their influence on the temperature and chemical-potential dependent pole masses, the possible signatures of a CEP and the restoration of chiral symmetry.

This paper is organized as follows: In \Sec{subsec:zuber} we discuss the formulation of massive vector fields and verify the extraction of the vector spectral function from the imaginary part of the resulting transverse retarded propagator. After motivating and introducing the effective model in \Sec{subsec:model} we show how the massive vector meson propagators can be implemented into the FRG framework in \Sec{subsec:frg_propagator}. In \Sec{sec:results} we then present the numerical results, first for the Euclidean parameters in \Sec{subsec:numerics_and_flow}, then the real-time results in the vacuum in \Sec{subsec:results_vacuum}, and finally the $\rho$ and $a_1$ spectral functions at finite temperature and chemical potential in \Sec{subsec:results_medium}. We conclude with our summary in \Sec{sec:summary}. An effective Lagrangian for massive vector fields based on anti-symmetric rank-2 tensors \cite{Gasser:1983yg} is shown to lead to vector correlators of the same form in \App{AppHodge}. Further technical details concerning the FRG flow equations and the analytic continuation procedure are provided in Appendices \ref{sec:flow_equations} and \ref{sec:analytic_continuation}.

\vspace{.5cm}

\section{Theoretical setup}\label{sec:setup}

\subsection{Massive vector fields and covariant time ordering}\label{subsec:zuber}

To describe massive vectors by fundamental fields in an effective theory is known to be problematic \cite{Itzykson1980,Nakanishi1990}, if not impossible without Higgs mechanism. In the Proca formalism the transversality of the corresponding Green functions is maintained only on-shell which, among other problems, leads to a pathological ultraviolet behavior. While this is fixed in the Stueckelberg formalism, one is then left with spurious massless single-particle contributions to the vector Green functions when restoring transversality in the Stueckelberg limit. Essentially the same is true for Nakanishi's $B$-field formalism.

There is of course no problem with massive vectors in Abelian-Higgs or Fradkin-Shenker models, for example, or in the Standard Model for that matter. However, the physical and hence gauge-invariant vectors are then necessarily described by composite fields \cite{Frohlich:1981yi,Maas:2019nso}. Here we adopt a somewhat simpler approach to describe fluctuations due to (axial-)vector mesons within our FRG framework below. It starts from the fairly general point of view, describing massive vectors as single-particle contributions, which may well be composites, to the corresponding conserved vector-current correlation functions. In order to understand their spectral representations, on the other hand, it is important to remember the subtlety in defining covariant time ordering for vector or higher-rank tensor field operators \cite{GROSS1969269,Watanabe,Treiman1972,Treiman1986}.

Our brief review here follows the discussion in \cite{Itzykson1980} for the simplest example of the correlation function of a conserved $U(1)$ current $j_\mu(x)$. As for any Feynman propagator, its causal Green function must be covariant while the naive time-ordered product is not. One therefore defines covariant time ordering,
\begin{align}
  & \hskip -.2cm \big\langle T_\mathrm{cov} \,j_\mu (x) j_\nu (0) \big\rangle
  = \\
 & \hskip .2cm  \theta(x^0) \, \big\langle j_\mu (x) j_\nu (0) \big\rangle \, + \,
\theta(-x^0)\, \big\langle j_\nu (0) j_\mu (x) \big\rangle \, + \, \tau_{\mu\nu}(x) \,, \nonumber
\end{align}
which differs from naive time ordering by a seagull term $\tau_{\mu\nu}(x)$ proportional to a delta distribution at $x=0$. Such a seagull term must occur whenever the corresponding equal-time commutators between different current components contain a Schwinger term  \cite{Schwinger_field_commutators}. In our example this is the case for
\begin{align}
  \big\langle  \big[j_0 (x), j_i (0)\big] \big\rangle \Big|_{x^0=0}=
  \mathrm i \partial_i \delta^3(\vec{x}) \int_0^\infty \!\! ds \, \frac{\rho(s)}{s}\, , \label{SchwingerT}
\end{align}
where $\rho(s) \ge 0$ is the spectral function of the current-current correlators. Together with the covariance of their causal Green functions, the requirement that Schwinger terms are canceled from Ward identities \cite{brown1967covariance} then fixes the seagull term uniquely, in the present case,
\begin{equation}
  \tau_{\mu\nu}(x) = \mathrm i  \big( g_{\mu 0} g_{\nu 0} - g_{\mu\nu} \big) \, \delta^4(x) \, \int_0^\infty \!\! ds \, \frac{\rho(s)}{s}\, ,  
\end{equation}
with metric and other conventions as in \cite{Itzykson1980}, where it is explicitly demonstrated that this leads to a spectral representation of the causal Green function, with covariant time ordering, 
\begin{align}\label{eq:propagator}
&\big\langle T_\mathrm{cov} \,j_\mu (x) j_\nu (0) \big\rangle =\\
  &\hspace{4mm}-\mathrm{i}\int_0^\infty \!\! ds \, \frac{\rho(s)}{s}\, 
\int \!\frac{d^4 p}{(2\pi)^4}\,\mathrm{e}^{-\mathrm{i}px}\; \frac{p^2 g_{\mu\nu}-p_\mu p_\nu}{p^2- s + \mathrm{i}\epsilon} \, . \nonumber 
\end{align}
This is manifestly transverse and covariant as it should be, with a measure given by the semi-positive spectral density $\rho(s)$. This is the spectral function of the conserved $U(1)$ current per charge squared.
Assuming a (minimal) first-order interaction $g_v \, j_\mu V^\mu$  of the current with a vector field $V_\mu(x)$, with coupling $g_v$, it is related to the vector field's spectral function $\rho_v(s)$ by 
\begin{equation}
  g_v^2 \rho(s) = s^2 \rho_v(s) \, .
\end{equation}
A massive single-particle contribution of strength $Z$ in $\rho_v(s) $, corresponding to a stable vector meson of mass $m_v$,  will therefore contribute to the current spectral function with a term
\begin{equation}
  \rho(s) = \frac{m_v^4}{g_v^2} \, Z \, \delta(s-m_v^2) \, + \, \dots\,.
\end{equation}
In order to describe such a possibly composite state by a
single-particle contribution to the vector-meson field $V_\mu$ in the spirit of vector-meson dominance, i.e.~with a current-field identity
\begin{equation}
j_\mu(x)  \, = \, \frac{m_v^2}{g_v}  \, V_\mu (x) \, , \label{CFI}
\end{equation}
we therefore need to arrive at a transverse vector-meson propagator $D^{V}_{\mu\nu} $ with a single-particle contribution of the form (here still in Minkowski space),
\begin{equation}
  D^V_{\mu\nu}(p)  \, = -\mathrm{i} \frac{Z}{m_v^2}   \; \frac{p^2 g_{\mu\nu}-p_\mu p_\nu}{p^2- m_v^2 + \mathrm{i}\epsilon} \, + \, \dots \, . \label{vectorprop}
\end{equation}
This is not of the form of a massive Proca propagator, and it differs by a factor $p^2/m_v^2$ from the transverse propagator that results in the Stueckelberg limit. In particular, it does therefore not come along with massless single-particle contributions. The price, however, is the poor ultraviolet behavior when viewed as the propagator of an elementary field. Although we have therefore obviously not succeeded to describe an off-shell vector meson by an elementary field, this form is still useful for our effective description of vector-meson fluctuations. The correct low-energy effective Lagrangian for a massive transverse propagator of this form, in fact, starts from describing left and right-handed vectors in terms of (anti-)self-dual field strengths \cite{Gasser:1983yg} which can then be re-expressed in terms of conserved 4-vectors to yield propagators of the form in Eq.~(\ref{vectorprop}) as we describe in \mbox{App. \ref{AppHodge}.}

The introduction of a regulator function $R_k(p)$ to suppress fluctuations of momentum modes $p<k$ in the FRG framework requires to modify Ward identities accordingly \cite{Litim:1998nf}.  We will therefore use an Ansatz for fluctuating vector mesons which contains additional longitudinal terms that vanish with $k\to 0$ in a way such that a transverse vector-meson propagator of the form as in Eq.~(\ref{vectorprop}) is obtained in the infrared as explained in Sec.~\ref{subsec:frg_propagator} below.  

In order to extract spectral functions from the results of integrating the analytically continued FRG flow equations we also need the imaginary parts of the retarded (axial-)vector propagators. This is slightly subtle for the same reasons, Schwinger and seagull terms, but the result will luckily be just as one would naively expect:

The spectral function $\rho_v(s) $ is originally defined from the commutator of the vector field. Via \Eq{CFI} this is essentially the same as that of the currents $j_\mu(x)$ which, however, includes the Schwinger term in \Eq{SchwingerT},
\[
\big\langle \big[ V_\mu(x), V_\nu(0) \big] \big\rangle = - \!\int_0^\infty\!\!\! ds \, \frac{s^2\rho_v(s)}{m_v^4} \Big( g_{\mu\nu} + \frac{\partial_\mu \partial_\nu}{s} \Big) \, \mathrm i \Delta(x;s) \, ,
\]
written in terms of the invariant delta function
\[ \mathrm i \Delta(x;m^2)   = \int \frac{d^4p}{(2\pi)^4} \, \epsilon(p_0)
\, 2\pi \delta (p^2 - m^2) \, \mathrm e^{-\mathrm i px}\, . \]
As usual, its Fourier transform therefore essentially defines the spectral function. In particular, we obtain here,
\begin{align}  
\int d^4x \, \mathrm e^{\mathrm i px}\, \big\langle \big[ V_\mu(x), V_\nu(0) \big] \big\rangle =& \\ 
&\hskip -2.2cm   -  2\pi \epsilon(p_0) \theta(p^2)  \, \frac{p^2 \rho_v(p^2)}{m_v^4} \Big( p^2 g_{\mu\nu} - p_\mu p_\nu \Big) \, . \nonumber 
\end{align}
Expressing the invariant delta function by the imaginary part of the retarded Green function, 
\[
\mathrm i \Delta(x;m^2)  = - 2\, \text{Im} \Delta_R(x;m^2) \, , \]
we can therefore write
\begin{align}
  &\epsilon(p_0) \theta(p^2)  \, p^2\rho_v(p^2) \Big( p^2 g_{\mu\nu} - p_\mu p_\nu \Big) = \\
  & \frac{1}{\pi} \int_0^\infty\!\! ds \, s^2\rho_v(s)
  \Big( g_{\mu\nu} - \frac{p_\mu p_\nu}{s} \Big) \, \text{Im} \frac{-1}{(p_0+\mathrm i\epsilon )^2 - \vec p^2 -s} \, .\nonumber
\end{align}
This is not quite the imaginary part of the retarded propagator corresponding to \Eq{eq:propagator} yet. However, because the imaginary part of the Fourier transform of $\Delta_R(x,m^2) $ has support only at $p^2 = m^2$ we can trade powers of $p^2$ for matching powers of $s$ in this spectral integral to write 
\begin{align}
  \epsilon(p_0) \theta(p^2)  \, \rho_v(p^2) \Big( g_{\mu\nu} - \frac{p_\mu p_\nu}{p^2} \Big) &= \\
  &\hskip -3cm  - \frac{1}{\pi} \, \text{Im} \int_0^\infty\!\! ds \, \frac{\rho_v(s)}{s} \, 
  \frac{p^2 g_{\mu\nu} - p_\mu p_\nu}{(p_0+\mathrm i\epsilon )^2 - \vec p^2 -s} \, .\nonumber
\end{align}
This confirms that we can safely extract also a vector spectral function from the discontinuity along the cut of the transversally projected vector propagator with spectral representation as in \Eq{eq:propagator}, i.e.~from the transverse imaginary part of a retarded vector propagator of the form,
\begin{equation}
  \label{transvecprop}
    D^{T,R}_{\mu\nu}(p)  \, =  \int_0^\infty \!\! ds \, \frac{\rho_v(s)}{s} \; \frac{p^2 g_{\mu\nu}-p_\mu p_\nu}{(p_0+\mathrm{i}\epsilon)^2- \vec p^2- s } \, .
\end{equation}


\vspace{.2cm}

\subsection{Gauged linear sigma model with quarks and the FRG}\label{subsec:model}
In this section we briefly introduce the effective model we employ. For a more detailed discussion of the model we refer to \cite{Rennecke:2015eba} and \cite{Jung:2016yxl}.

Starting point is the linear sigma model with quarks, often used as effective low-energy model for two-flavor QCD to study the chiral phase transition. It contains the isotriplet $\vec{\pi}$ and the isosinglet $\sigma$ as chiral partners in the scalar sector which are coupled to quark-antiquark fields with a Yukawa-type interaction. In these type of models one has a CEP at low temperature and large quark chemical potential, separating a crossover transition at larger temperatures from a first order phase transition at lower ones, into a dense region of self-bound quark matter. The $\sigma$ field contains an exactly massless critical mode at the CEP. Its expectation value $\sigma_0$ serves as an order parameter for (spontaneous) chiral symmetry breaking.

In such models, vector mesons are usually introduced as the gauge fields of a local flavor symmetry, first proposed by Sakurai \cite{1960AnPhy} and later extended to the full chiral group to introduce the vector and axial-vector isotriplets $\vec{\rho}$ and $\vec{a}_1$ as the gauge fields of a local chiral symmetry $SU(2)_L\times SU(2)_R$ \cite{Lee:1967ug}. This idea of local gauge invariance was also the origin of the concept of vector-meson dominance (VMD) \cite{Schildknecht:2005xr}. Alternatively, one sometimes imposes only a global chiral symmetry rather than a local one \cite{Urban:2001ru,Eser:2015pka}.

In this work we employ the gauged linear sigma model with quarks based on the assumption of VMD within the framework of the functional renormalization group. The idea of the FRG is to introduce a momentum scale $k$ and to integrate out fluctuations with momenta larger than this scale. To this end one defines the so-called effective average action $\Gamma_k$ which interpolates between the classical action $S\equiv \Gamma_{k\rightarrow \Lambda}$ at some chosen ultraviolet (UV) cutoff scale $\Lambda$ and the full quantum effective action $\Gamma_{k\rightarrow 0}$ at the infrared (IR) scale $k\rightarrow 0$. Technically this is archived by implementing a regulator function $R_k$ which suppresses low-momentum fluctuations. The change of $\Gamma_k$ when changing the scale $k$ is described by the so-called Wetterich equation \cite{Wetterich:1992yh},
\begin{align}
\label{eq:wetterich}
\partial_k \Gamma_k = \frac{1}{2} \Tr \left[\partial_k R_k^{\phi}~ \left(\Gamma^{(2)}_k[\phi]+R_k^{\phi}\right)^{-1}\right]\,.
\end{align}

One hence starts with an Ansatz for the effective average action $\Gamma_{k=\Lambda}$ at the UV scale and integrates out fluctuations momentum shell by momentum shell until arriving at the IR scale $\Gamma_{k\rightarrow 0}$. From suitable functional derivatives of \Eq{eq:wetterich} one furthermore obtains flow equations for the corresponding scale dependent $n$-point vertex functions $\Gamma^{(n)}_k$. 

\begin{figure*}[t]
	\includegraphics[width=0.9\textwidth]{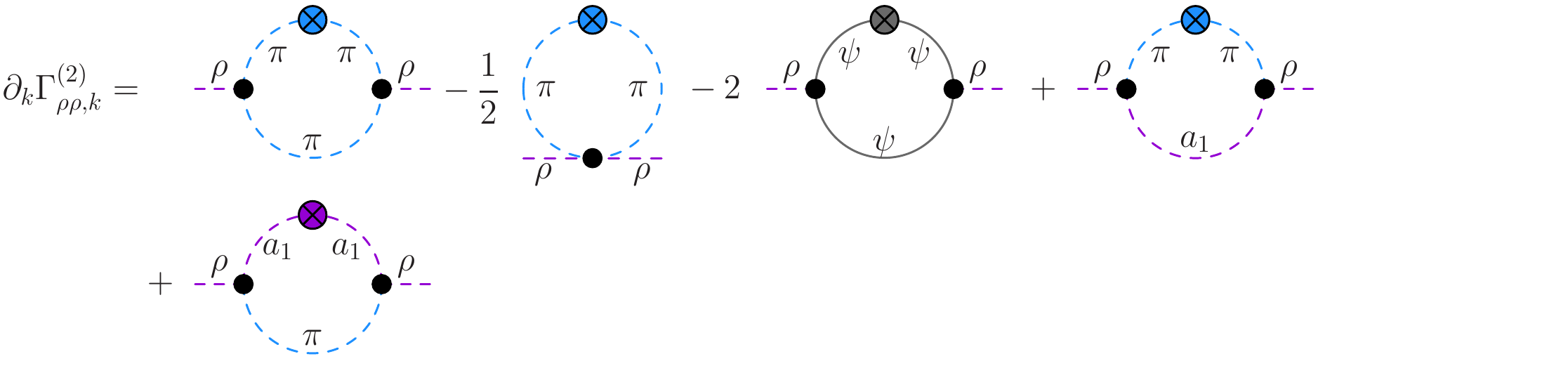}\\
	\vspace{3mm}\hspace{-3.5mm}
	\includegraphics[width=0.9\textwidth]{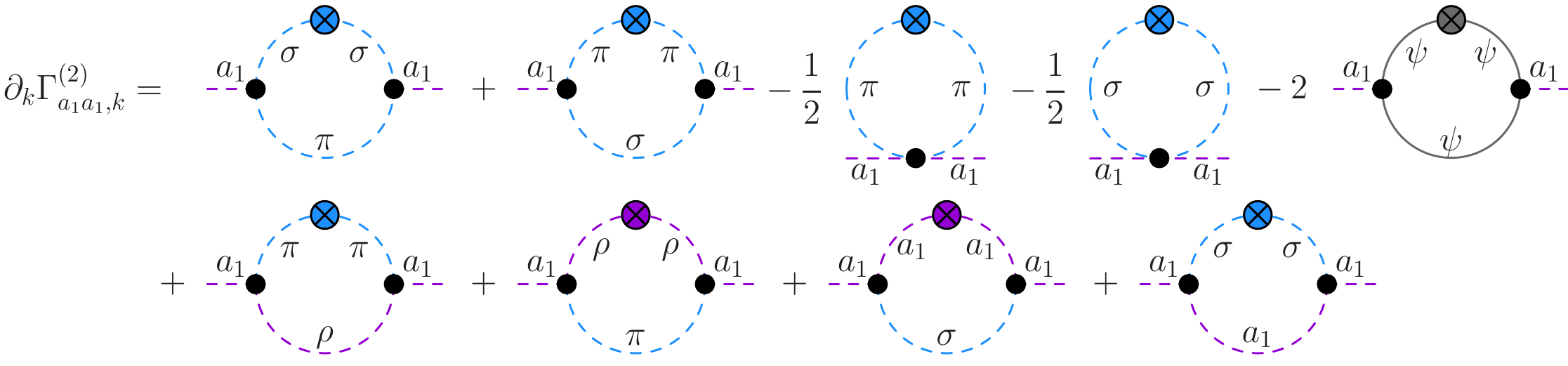}
	\caption{(color online) Flow equations for the $\rho$ and $a_1$ two-point functions in diagrammatic form. Vertices are indicated by black filled dots, regulator insertions by crossed circles. The color of the lines and regulators represents the type of field: blue for scalar and pseudo-scalar mesons, black for fermions and purple for vector mesons.}\label{fig:flow_equations_two-point}
\end{figure*}

A widely used Ansatz for the Euclidean effective average action is provided by the leading order in a derivative expansion, in the literature also referred to as the local potential approximation (LPA). For the extended linear-sigma model with gauged $SU(2) \times SU(2) \simeq SO(4) $ chiral symmetry, for example, this Ansatz is of the following form,
\begin{widetext}
\begin{align}\label{eq:lagrangian}
\Gamma_k = \int d^4x &\bigg\{\bar{\psi} \left(\slashed{\partial}-\mu\gamma_0+
h_{s}\left(\sigma +\mathrm{i} \vec{\tau}\dt\vec{\pi}\gamma_5\right)+
\mathrm{i} h_{v} \left(\gamma_{\mu} \vec{\tau}\dt\vec{\rho}_{\mu}+\gamma_{\mu}\gamma_5 \vec{\tau}\dt\vec{a}_{1\mu}\right)
\right)\psi+ U_{k}(\phi^2)-c\sigma\nonumber\\
&+\frac{1}{2} (D_\mu \phi)^\dagger D_\mu \phi 
+\frac{1}{8} \tr\big(V_{\mu\nu} V_{\mu\nu}\big)
+\frac{1}{4}m_{v,k}^2 \, \tr \big(V_{\mu}V_{\mu}\big)
+\frac{1}{4}\lambda_{k}\, \tr \big(\partial_{\mu}V_{\mu}\big)^2
\bigg\}\, ,
\end{align}
\end{widetext}
where, compared to \cite{Jung:2016yxl}, we only added the last term with scale-dependent Stueckelberg parameter $\lambda_k$ for now. We will have to go beyond this leading order in the derivative expansion to accommodate propagators of the form in Eq.~(\ref{vectorprop}) for massive vector and axial-vector fields, however. This will be discussed in the next subsection. 


As usual, the (pseudo-)scalar mesons are collected in 4-vectors, and the (axial-)vector mesons in skew-symmetric hermitian matrices transforming in the adjoint $SO(4)$ representation,
\begin{align}
&\phi\equiv (\vec{\pi},\sigma)^T\,,\\
&V_\mu\equiv \vec{\rho}_{\mu}\dt\vec{T}+\vec{a}_{1\mu}\dt\vec{T}^5\,.
\end{align}
The six hermitian generators  $\vec{T}$ and $\vec{T}^5$ in the $\mathfrak{so}(4)$ algebra are defined such that $T_i^L\equiv\frac{1}{2}(T_i-T_i^5)$ and $T_i^R\equiv\frac{1}{2}(T_i+T_i^5)$ generate the $(1,0)$ and $(0,1)$ representations of $SU(2)_L \times SU(2)_R$,
\begin{align}
[T_i^L,T_j^L] &= \mathrm{i}\epsilon_{ijk}T_k^L\,,\\
[T_i^L,T_j^R] &= 0\,,\\
[T_i^R,T_j^R] &= \mathrm{i}\epsilon_{ijk}T_k^R\,.
\end{align}
The first part in \Eq{eq:lagrangian} describes two quark flavors with chemical potential $\mu$ plus Yukawa couplings $h_{S}$ to the (pseudo-)scalar mesons and $h_{v}$ to the (axial-)vectors.
Then we have the $O(4)$-invariant effective potential including scalar meson self-interactions and the explicit chiral-symmetry breaking term $c\sigma$. Next we have the kinetic term for the scalars coupled minimally, via a covariant derivative $\partial_\mu + \mathrm i g V_\mu$, with gauge coupling $g$, to the vector mesons $V_\mu$ whose non-Abelian field-strengths tensor would be given by 
\begin{align}
V_{\mu\nu} = \partial_\mu V_\nu-\partial_\nu V_\mu+\mathrm{i}g\left[V_\mu,V_\nu\right] \, .
\end{align}
As in \cite{Jung:2016yxl} we will neglect their self-interactions here, however, and only maintain the Abelian part. Then, the last three terms on the right hand side in \Eq{eq:lagrangian} together represent one equal Abelian Stueckelberg Lagrangian for each of the six massive vector fields described by $V_\mu$, related to one another by a global $SO(4)$ symmetry. This part will need further amendment as described below.

Moreover note that in a description with a fully gauged  $SU(2) \times SU(2) $ local chiral symmetry and a photon coupling according to Kroll, Lee and Zumino \cite{Kroll:1967it} one would have to require $h_v= g/2$ in order to be consistent with the original VMD interaction with coupling $g_v=g$. For a most direct comparison we instead use $h_{s}=h_{v}$ here, as was done in the previous study in Ref.~\cite{Jung:2016yxl}. 

Taking two functional derivatives of the Wetterich equation, \Eq{eq:wetterich}, and using an Ansatz of the from in \Eq{eq:lagrangian} one obtains flow equations for the Euclidean $\rho$ and $a_1$ meson two-point functions. Without truncations this would still result in an infinite tower of equations as the flow of $\Gamma^{(n)}_k$ involve vertex functions up to $\Gamma^{(n+2)}_k$. To obtain a closed set of equations we therefore extract the $n$-point functions needed inside the flow equation for $\Gamma^{(2)}_k$
from the LPA Ansatz of the effective action in \Eq{eq:lagrangian}. This thermodynamically consistent and symmetry preserving scheme is self-consistent for the 2-point functions only in the limit of vanishing momenta, however. Moreover, the scale-dependent three and 4-point functions $\Gamma^{(3)}_k$ and $\Gamma^{(4)}_k$ are themselves momentum independent in this truncation. Self-consistent 2-point functions at finite momenta, and three and 4-point functions with non-trivial 
internal substructur will be left as important extensions for future studies. The resulting equations are shown diagrammatically in \Fig{fig:flow_equations_two-point}, where one can already identify the kind of processes that can occur in this truncation. After the analytic continuation procedure, their real and imaginary parts basically determine the spectral functions, cf. \App{sec:analytic_continuation}.

\subsection{Vector-meson fluctuations}\label{subsec:frg_propagator}

In this section we explain how to implement propagators for single-particle contributions of the form in Eq.~(\ref{vectorprop}) from massive (axial-)vectors, as discussed in \Sec{subsec:zuber}, into the FRG flow equations. Since these are formulated in Euclidean spacetime, we first write the corresponding Feynman propagator in \Eq{vectorprop} as transverse Euclidean momentum-space propagator,
\begin{align}\label{eq:eucl_zuber}
D_{\mu\nu}^{T,\text{E}}(p)= \frac{Z}{m^2}\frac{-p^2}{p^2+m^2}\, \Pi^{T}_{\mu\nu}(p)\,,
\end{align}
with transverse projector $\Pi^{T}_{\mu\nu}(p)=\delta_{\mu\nu}-p_\mu p_\nu/{p^2}$. 

We also need the corresponding two-point function $\Gamma_{\mu\nu}^{(2),\text{E}}(p)$ in the FRG calculation, given by the inverse of the propagator. One therefore adds a constant longitudinal part in Stueckelberg formalism, as also done for example in \cite{Urban:2001ru,Eser:2015pka}. The Stueckelberg Lagrangian in \Eq{eq:lagrangian} corresponds to the following LPA-like Ansatz for the massive vector two-point functions
\begin{align} \label{stueckelberg-LPA}
  \Gamma_{\mu\nu,k}^{(2),\text{E}}(p) =& \,(p^2+m_{v,k}^2)\,\Pi^{T}_{\mu\nu}(p) \\
         & \hskip .8cm +\,( \lambda_k \,p^2+ m_{v,k}^2)\,\Pi^{L}_{\mu\nu}(p)\,, \nonumber
\end{align}
where $ \Pi^{L}_{\mu\nu}(p) = \delta_{\mu\nu} - \Pi^{T}_{\mu\nu}(p)= p_\mu p_\nu/{p^2}$.

With such an approach, based on the Stueckelberg LPA Ansatz (\Eq{eq:lagrangian}), however, one is always left with non-vanishing unphysical longitudinal contributions of the vector meson propagators inside the loops (see \Fig{fig:flow_equations_two-point}). In particular, these unphysical fluctuations can lead to sizable negative contributions to spectral functions and hence to positivity violations, indicating that even the Abelian gauge invariance is lost. In order to fix this problem, we modify the massive vector part of our Stueckelberg LPA Ansatz for the scale-dependent effective average action in the following way: In the parts that are quadratic in the vector fields, in momentum space, we first replace the Stueckelberg LPA form of $ \Gamma_{\mu\nu,k}^{(2),\text{E}}$ in \Eq{stueckelberg-LPA} by 
\begin{align} \label{beyond-LPA-vectors}
  \Gamma_{\mu\nu,k}^{(2),\text{E}}(p) =& \,Z_k^T(p^2)\, (p^2+m_{v,k}^2)\,\Pi^{T}_{\mu\nu}(p) \\
         & \hskip .8cm +\, \lambda_k Z_k^L(p^2)  \,( p^2+  m_{v,k}^2/\lambda_k)\,\Pi^{L}_{\mu\nu}(p)\,, \nonumber
\end{align}
with momentum dependent longitudinal and transverse wavefunction renormalizations $Z_k^L(p^2)$  and $ Z_k^T(p^2)$ in addition to the scale dependent mass and Stueckelberg parameters. In order to restore the transversality of the corresponding propagator in the infrared, 
\begin{align}\label{eq:mWI}
p_\mu D^{\mu \nu}_{k\rightarrow 0}=0 \, ,
\end{align}
which is anyway the proper generalization of the corresponding Ward identity, once a regulator function $R_k(p^2)$ is introduced \cite{Litim:1998nf},  we then choose a running Stueckelberg parameter that starts at the ultraviolet cutoff $k=\Lambda$ with some finite value $\lambda_\Lambda > 0  $ and tends to zero for $k\to 0$ in order to effectively  make the longitudinal fluctuations, now with mass $m^2_{v,k}/\lambda_k$,  infinitely heavy in the infrared. One simple choice that does the job is
\begin{equation}
\lambda_k \, = \, \frac{k^2}{\Lambda^2} \, ,\quad \mbox{such that} \;\;  \lambda_\Lambda = 1\, ,\; \lambda_k \stackrel{k\to 0}{\longrightarrow} 0\,  . \label{eq:lambdadef}
\end{equation}
In addition,  we require the longitudinal renormalization function to cancel 
the explict factor $\lambda_k$ in front of the longitudinal term. A simple way to realize this is to set 
\begin{equation}
  \lambda_k Z_k^L(p^2) = Z_k^T(p^2)
\end{equation}
which we assume to be independent of the running Stueckelberg parameter $\lambda_k$. Note that assuming $Z^L_k = Z^T_k $ and both to be independent of $\lambda_k$ instead, would essentially lead back to the Proca propagator for $\lambda_k \to 0$ in the infrared (i.e.~up to the factor $Z^T_k $). In the present setup on the other hand, longitudinal and transverse components start out equally at the ultraviolet cutoff scale for $k=\Lambda$ so that the two-point function is altogether proportional to $\delta_{\mu\nu}$ (as in Feynman gauge). Relative to the transverse mass, the longitudinal mass becomes heavier and heavier and the unphysical longitudinal fluctuations switch themselves off automatically during the flow.     

Finally, for the transverse part to correctly model the single-particle contribution to the vector correlator (\ref{vectorprop}), cf.~\Eq{eq:eucl_zuber} with scale dependent mass $m_{v,k}$ and strength $Z_k$, all we have now left to do, is to set
\begin{equation}
 Z^T_k(p^2) = - Z_k^{-1} m^2_{v,k}/p^2 \equiv - m^2_{0,k}/p^2 \, , \label{eq:ZTdef}
\end{equation}
with an independent mass parameter $m_{0,k}^2 = m^2_{v,k}/Z_k$ which for $Z_k < 1$ we expect to be larger than the scale dependent (pole-)mass of the vecor meson $m_{v,k}^2$,  in particular for $k\to 0$ in the infrared. We will start the flow with the boundary condition at the ultraviolet cutoff $k=\Lambda $ with $Z_\Lambda = 1$, corresponding to the full spectral strength initially contained in this single-particle contribution.

Note that the somewhat ambiguous details in the treatment of the unphysical longitudinal fluctuations are irrelevant here: Just as the vector-meson mass parameter, the longitudinal mass starts out at a rather large initial value of about
\begin{align}
m_{l,\Lambda} =m_{v,\Lambda} \,  \approx 1 \, \mbox{GeV} 
\end{align}
as compared to a UV cutoff for which we typically use $\Lambda = 1.5$~GeV. Because the longitudinal mass 
\begin{align}
m_{l,k} = \frac{\Lambda}{k}\, m_{v,k} 
\end{align}
increases rapidly with lower $k$ from there on, these longitudinal fluctuations are strongly suppressed during the flow.
In principle, their suppression can be further controlled by the initial value of the Stueckelberg parameter $\lambda_\Lambda$. One then verifies that the results are in fact independent of this parameter for sufficiently small $\lambda_\Lambda$. Our choice of $\lambda_\Lambda = 1$ seems rather natural but is by no means mandatory, it simply turns out to be sufficiently small for the parameters used here at least at low temperatures. For higher temperatures, e.g.~for the results presented in the next section with $ T = 100$~MeV and above, we in fact observe that longitudinal fluctuations can occasionally still produce small spurious contributions to capture processes during the flow, when their initial mass is not large enough for $\lambda_\Lambda = 1$. In such cases we simply reduce  $\lambda_\Lambda $ further, until we observe no noticeable dependence on $\lambda_\Lambda$ any more. 

In contrast, we have been very careful in modeling the transverse fluctuations to correctly describe the single-particle contributions to the full (axial-)vector correlators, including momentum and field independent wavefunction renormalization in form of the scale-dependent strength $Z_k$. This treatment of the (axial-)vector fluctuations thus in this sense parallels what has been called the LPA' truncation for the (pseudo-)scalar sector in the literature.    

Choosing a transverse and longitudinal regulator functions of the same form,
\begin{equation}
  R_{\mu\nu,k}^{T,L}(p) = Z_k^{T,L}(p^2) \, k^2 \, r_k(p) \, \Pi^{T,L}_{\mu\nu}(p) \, ,
\end{equation}
with a suitable dimensionless regulator function $r_k(p)$,
and using~(\ref{eq:lambdadef})--(\ref{eq:ZTdef}) in \Eq{beyond-LPA-vectors}, the Ansatz for the scale-depended vector propagator with regulator becomes,
\begin{align}\label{eq:frg_zuber}
  D_{\mu\nu,k}^{\text{E}} (p) \equiv& \, \Big(\Gamma_{k}^{(2),\text{E}}(p) + R_{k}(p) \Big)_{\mu\nu}^{-1}\\ 
  =&\, 
  \frac{-p^2}{m_{0,k}^2 \,(p^2+  k^2 \,r_k(p)+ m_{v,k}^2  )}\,\Pi_{\mu \nu}^{T}(p) \nonumber\\
  &\hskip .4 cm 
  - \frac{p^2}{m_{0,k}^2\, (p^2+  k^2 \,r_k(p) +\frac{\Lambda^2}{k^2}m_{v,k}^2) }\,\Pi_{\mu \nu}^{L}(p) \, . \nonumber
\end{align}
For $k\to 0$ we recover \Eq{eq:mWI}, and the transverse part reduces to the desired single-particle contribution of a massive vector state as in \Eq{eq:eucl_zuber}. The flow of the mass parameter $m_{0,k}^2 = m_{v,k}^2/Z_k$ is in general different from that of the LPA' vector-meson mass $m_{v,k}^2$, of course. As mentioned above, we will assume their initial values to be the same at the UV cutoff, with $Z_\Lambda = 1$, i.e.
\begin{align}
m_{0,\Lambda}^2=m_{v,\Lambda}^2 \, .
\end{align}
Due to the fluctuations in the interacting theory one then expects  $Z_k < 1$   and, with $\lambda_k = k^{2}/\Lambda^2$, therefore the overall ordering
\begin{align}
m^2_{l,k} > m_{0,k}^2 > m_{v,k}^2 \, , \quad\mbox{for}\;\;  k <\Lambda \,.
\end{align}
This behaviour is confirmed explicitly in the numerical calculations as described in the next section,  \Sec{subsec:numerics_and_flow}.

As mentioned above, the LPA' Ansatz in \Eq{eq:frg_zuber} is used to describe the fluctuations due to the single-particle contributions of massive vectors on the right hand side of the FRG flow equations in Fig.~\ref{fig:flow_equations_two-point}, in our present truncation. The result of the integrated flow on the left hand side yields the corresponding full two-point functions including widths and various thresholds whose transverse parts will have a spectral representation as in \Eq{transvecprop}.  In a fully self-consistent calculation one would have to feed these back into the flow equations, recompute and iterate until convergence \cite{Strodthoff:2016pxx}. As in our previous studies \cite{Tripolt2014,Tripolt2014a,Jung:2016yxl} here we perform the first step in such an approach, thus neglecting the fluctuations due to the continuous contributions from the two-point functions inside the flow.

\vspace{.2cm}

\section{Numerical Results}\label{sec:results}

\subsection{Euclidean FRG flow and mass paramters}\label{subsec:numerics_and_flow}

In this section we discuss the numerical procedure for solving the Euclidean flow equations and the resulting flow of the Euclidean (curvature) mass parameters. Since the setting used here in wide parts parallels that of Ref.~\cite{Jung:2016yxl} we keep the discussion brief and refer to this reference as well as our Appendix \ref{sec:flow_equations} for further details.

We start with solving the flow equation for the effective potential $U_k(\phi^2)$ which contains scalars and pseudo-scalars as well as quarks and antiquarks as fluctuating fields. Because the Euclidean mass parameters of the (axial-)vector mesons are relatively heavy, of the order of the UV cutoff \mbox{$\Lambda=1500$ MeV}, they are expected to contribute very little to the Euclidean flow of the effective potential and are therefore neglected. The flow equation for $U_k(\phi^2)$ is then solved with standard procedures by discretizing its argument in field space, and using the following simple form of the linear-sigma model in the symmetric phase at the UV scale $\Lambda$ as initial condition,
\begin{align}
U_{k=\Lambda}(\phi_i^2) = b_1 \,\phi_i^2+b_2\,
\phi_i^4\,.
\end{align}

Storing the $k$-dependent effective potential, we solve the flow equations for the vector-meson mass parameters $m_{v,k}^2$ and $m_{0,k}^2$ next, see \App{sec:flow_equations} for more details. With the parameters listed in 
\Tab{tab:parameters} (Set 1), we obtain following values for the chiral order parameter $\sigma_0$ and the Euclidean mass parameters at the IR scale of $k=40$~MeV,
\begin{alignat}{4}\label{eq:masses_values}
&\sigma_0 &&=93.0 \text{ MeV}\,, \quad &&m_{\sigma} &&=557.1 \text{ MeV}\,,\nonumber\\
&m_{\pi} &&=140.4 \text{ MeV}\,, \quad &&m_{\psi} &&=300.0 \text{ MeV}\,,\nonumber\\
&m_{\rho} &&=868.1 \text{ MeV}\,, \quad &&m_{a_1} &&=1363.1 \text{ MeV}\,,\nonumber\\
&m_{0} &&=1294.3 \text{ MeV}\,.
\end{alignat}
Note that the mesonic mass parameters here are not directly the physical meson masses. They are determined from the zero-momentum limit of the respective Euclidean two-point functions $\Gamma^{(2),E}(p)$. For the (pseudo-) scalars they agree with the corresponding curvatures in the $\pi$ and $\sigma$ directions of the effective potential in our thermodynamically consistent and symmetry-preserving truncation scheme.

\begin{table}[t]
	\centering
	\begin{tabular}{>{\vspace{-.2cm}\center} m{.9cm}|C{1cm}|C{0.5cm}|C{1.8cm}|C{1.2cm}|C{0.6cm}|C{1.4cm}}
        Set \# &  $b_1 \text{~[}\Lambda^2\text{]}$ & $b_2$ & $c\text{~[}\Lambda^3\text{]}$ & $h_s=h_v$&$g$ & $m_{v,\Lambda}\text{~[}\Lambda\text{]}$\\
	  \hline\hline
          1 &   0.381  & 0.2 & 0.5401$\cdot 10^{-3}$ &3.226 & 11.3 & 0.7067 \\
          2 &&&&& 11.8  & 0.684 \\
          3 &&&&& 10.5  & 0.74
	\end{tabular}
	\caption{Different UV parameter sets resulting in roughly the same pole masses, close to the physical ones of $\rho$ and $a_1$, all with the same quark-meson model parameters.}
	\label{tab:parameters} 
\end{table}

The analogous Euclidean masses of vector and axial-vector meson, on the other hand, then essentially result from tuning their coupling $g$ and UV mass parameter $m_{v,\Lambda} $ so that the corresponding pole masses $m_\rho^p$ and $m_{a_1}^p$  assume the approximate mass values of the physical $\rho(770)$ and the $a_1(1260)$ mesons. Since the $\rho$ and $a_1$ are resonances, we estimate their pole masses from the zeros of the real parts of the respective retarded two-point functions,
$ \Gamma^{(2),R}_{\rho/a_1}(p) $, 
which is a fairly good approximation as long as the widths of the resonances, i.e.~the imaginary parts of $\Gamma^{(2),R}_{\rho/a_1}$ in the resonance region, are sufficiently small. In our present qualitative study we are content with this  approximation and obtain the pole masses listed in \Tab{tab:masses} which are all reasonably close to the physical masses of $\rho$ and $a_1$ 
 for our representative UV parameters of Table \ref{tab:parameters}. For a more precise determination of masses and widths one would have to study the analytic structure of $\Gamma^{(2),R}_{\rho/a_1}$ and look for the resonance poles on the unphysical second Riemann sheet, see for example Refs.~\cite{Hidaka2003,Tripolt:2016cya}.

\begin{table}[t]
	\centering
	\begin{tabular}{>{\vspace{-.2cm}\center} m{1cm}|C{2cm}|C{2cm}}
        Set \# &  $ m_\rho^p \text{~[MeV]}$ & $m_{a_1}^p\text{~[MeV]}$ \\
	  \hline\hline
          1 & 776.3  & 1242.6 \\
          2 & 774.9  & 1266.2 \\
          3 & 770.2  & 1258.6 \\
          PDG & 775.26$\pm $0.25 & 1230$\pm $40 
	\end{tabular}
	\caption{Pole masses of (axial-)vector mesons for the 
          parameter sets of Table \ref{tab:parameters} compared to the estimates for $\rho(770)$ and $a_1(1260)$ from the Review of Particle Properties (2019).} 
	\label{tab:masses} 
\end{table}

\begin{figure}[b]
	\includegraphics[width=0.49\textwidth]{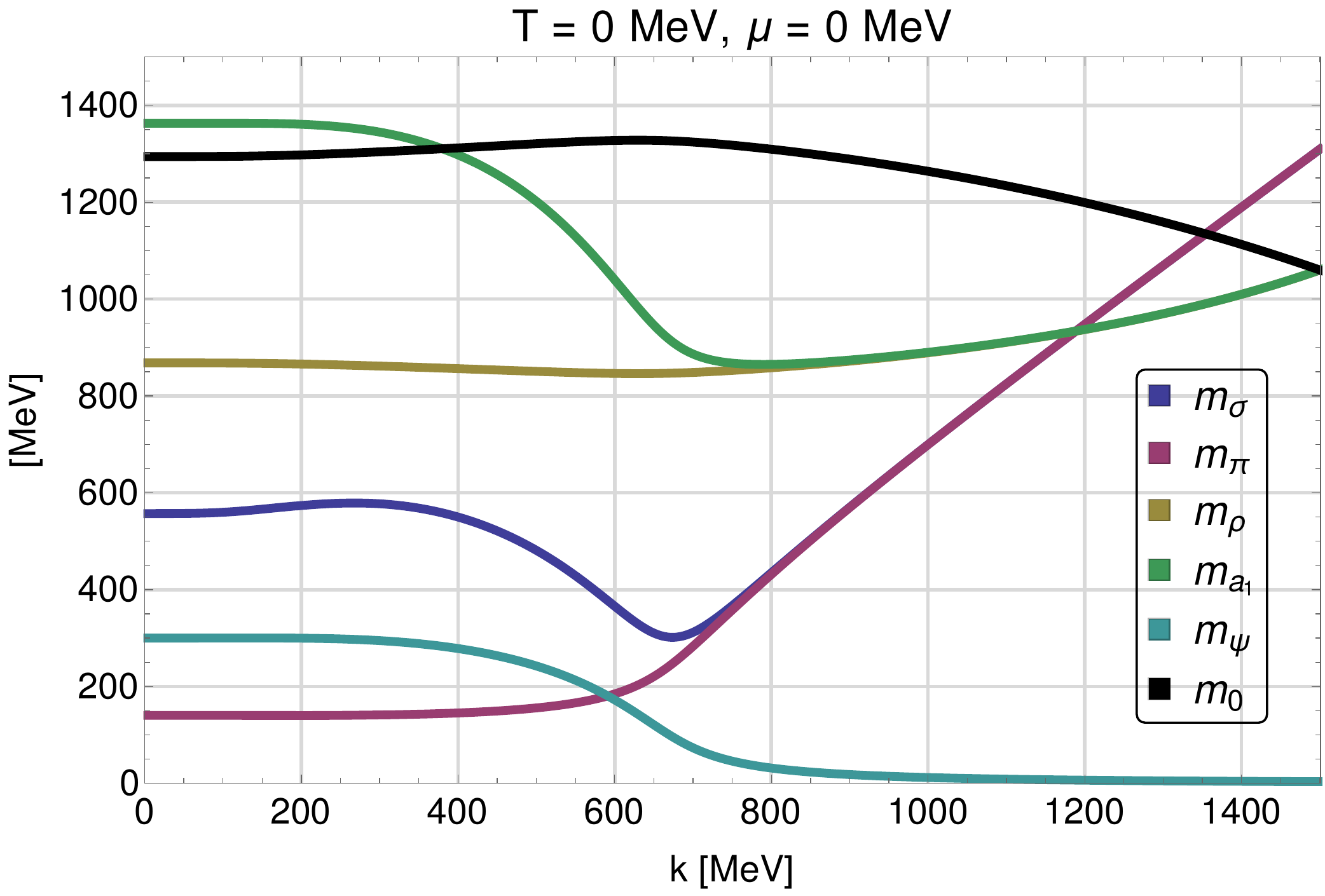}
        \caption{(color online) Flow of the Euclidean mass parameters
          in (\ref{eq:masses_values}) with the RG scale $k$ in the vacuum (parameter Set 1).}\label{fig:masses} 
\end{figure}

 The $k$-flow of the Euclidean masses and of the mass parameter $m_{0,k}$, all evaluated at the $k$-dependent minimum $\sigma_{0,k}^2$ of $U_k(\phi^2)$, is plotted in \Fig{fig:masses}. Starting at the UV scale where chiral symmetry is restored, the masses of the chiral partners $\pi$-$\sigma$ and $\rho$-$a_1$ are degenerate, the quark mass has its very small bare value. Taking fluctuations into account by successively lowering the scale $k$, the mass parameter $m_{0,k}$ immediately splits from its counterpart $m_{v,k}$ because their flow is not independent but in fact opposite in sign. In particular, one has
 \begin{equation}
   \frac{\partial_k m_{0,k}^2}{m_{0,k}^2} = - \frac{\partial_k m_{v,k}^2}{m_{v,k}^2} \, , 
 \end{equation}
cf.~App.~\ref{sec:flow_equations}. Moreover, the relation $m_{0,k}^2\geq m_{v,k}^2$ discussed in \Sec{subsec:frg_propagator} with $m_{0,\Lambda}^2 = m_{v,\Lambda}^2$ therefore holds by construction whenever the flow of the vector-meson mass parameter $m_{v,k}$ is predominantly negative as in \Fig{fig:masses}.

Lowering the scale further, spontaneous chiral symmetry breaking sets in, giving rise to an increase of the chiral order parameter $\sigma_{0,k}$. The quark acquires its dynamically generated constituent mass, and the masses of the chiral partners split up. At the IR scale we then arrive at the values listed in \Eq{eq:masses_values}.

\begin{figure}[b]
	\includegraphics[width=0.49\textwidth]{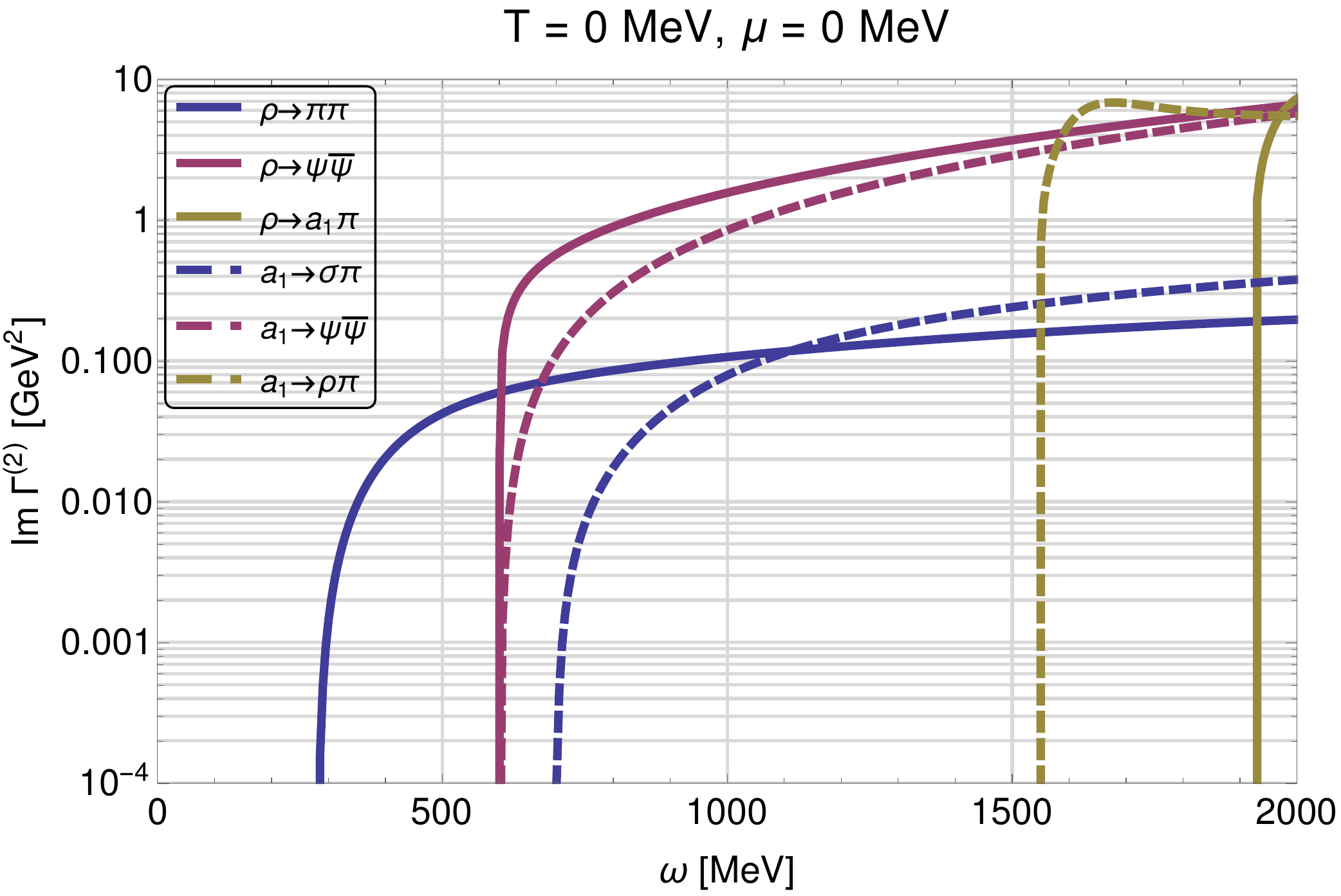}
	\caption{(color online) Imaginary parts of the retarded two-point functions of the $\rho$ (solid lines) and $a_1$ meson (dashed lines) for every process separately as a function of external frequency $\omega$ (evaluated at the fixed infrard minimum $\sigma = \sigma_0 $ using the parameters of Set 2).}\label{fig:imag} 
\end{figure}


\subsection{Spectral functions in the vacuum}\label{subsec:results_vacuum}

In this subsection we now explicitly demonstrate that the formalism presented in \Sec{subsec:frg_propagator} yields positive and physically meaningful contributions to the vector and axial-vector meson spectral functions.

Based on the Euclidean $k$-flow as input, from which all scale-dependent parameters are determined, we employ our standard analytic continuation procedure as outlined in \App{sec:analytic_continuation} to obtain the aFRG flow equations for the retarded two-point functions $\Gamma^{(2),R}_{\rho/a_1}(\omega)$, here at vanishing spatial momentum $\vec p$, for the $\rho$ and $a_1$ mesons from the diagrams shown in \Fig{fig:flow_equations_two-point}.

The spectral function as a function of the external frequency $\omega$ at vanishing external spatial momentum \mbox{$\left | \vec{p} \right|=0$ MeV} is then given by the discontinuity along the cut as verified explicitly for the transversally projected vector propagator at the end of Sec.~\ref{subsec:zuber}, which can be extracted from the real and imaginary parts of  $\Gamma^{(2),R}_{\rho/a_1}(\omega)$ as usual by
\begin{align}
\rho(\omega)=\frac{1}{\pi}\frac{\text{Im}\,\Gamma^{(2),R}(\omega)}{\left(\text{Re}\,
	\Gamma^{(2),R}(\omega)\right)^2+\left(\text{Im}\,\Gamma^{(2),R}(\omega)\right)^2}\,.
\end{align}

In \Fig{fig:imag} we show the imaginary parts of $\Gamma^{(2),R}_{\rho/a_1}(\omega)$ for every process separately. Since there are no in-medium capture processes in the vacuum, all thresholds indicate decays of off-shell (axial-)vector mesons $\rho$ and $a_1$ into the particle pairs as labeled in the figure, which can also be inferred from \Fig{fig:flow_equations_two-point}. In the present truncation, with only the single-particle contributions on the right hand side of the aFRG flow equations, the positions of the thresholds are still determined by this input, i.e.~by the Euclidean mass parameters at the IR scale. In a selfconsistent solution \cite{Strodthoff:2016pxx} they should eventually also be determined by the resulting physical pole masses, of course, possibly smeared out for resonances.

In the spirit of the grid-code technique, the $k$-dependent input for the aFRG flow equations for real and imaginary parts of $\Gamma^{(2),R}(\omega)$ 
is thereby normally evaluated at the fixed value of the $\sigma$-field variable that corresponds to the minimum of the effective potential at $\sigma = \sigma_{0} = 93$~MeV in the infrared. For the (axial-)vector mesons this results in mass parameters $m_\rho$ and $m_{a_1} $ that are considerably larger than those in \Eq{eq:masses_values} obtained from the $k$-dependent minimum $\sigma_{0,k}$ which further enhances the deviations of the two-particle thresholds from their physically expected values.

\begin{figure}[t]
	\includegraphics[width=0.49\textwidth]{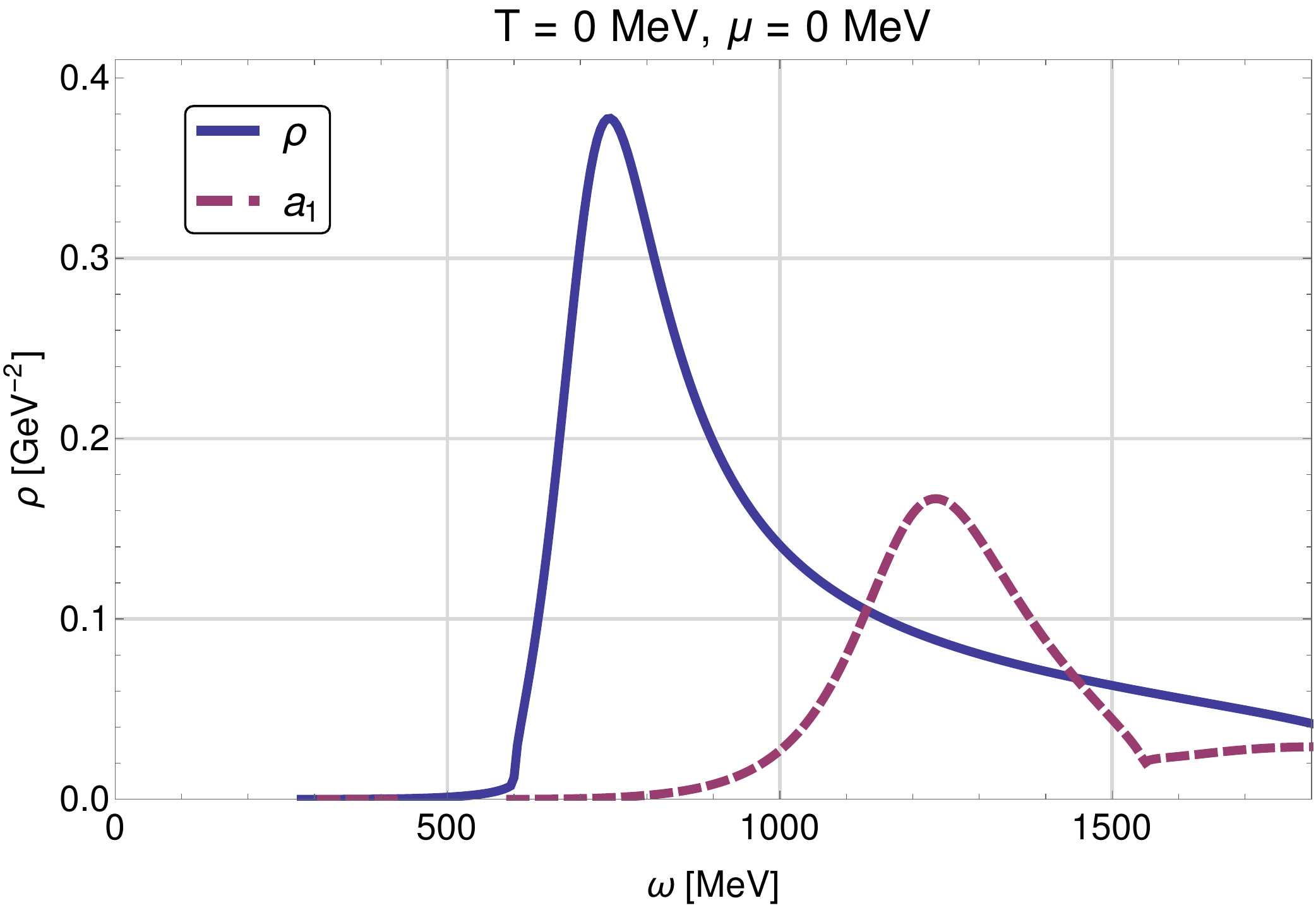}
	\caption{(color online) Vacuum spectral function of the $\rho$ (solid blue) and $a_1$ meson (dashed red) as a function of external frequency $\omega$ (evaluated at fixed $\sigma = \sigma_0 $, parameters of Set 2).}\label{fig:sf_linear} 
\end{figure}

In the present setup, for example, an off-shell $\rho$ meson can decay into an $a_1 + \pi$ pair, if the energy constraint is fulfilled, i.e.~for \mbox{$\omega \geq m_{a_1}+m_\pi \approx 1925.7$ MeV}, here with a large mass parameter \mbox{$m_{a_1} = 1785.3$ MeV}. Analogously, the decay $a_1\to \rho + \pi$ starts $\omega \geq m_{\rho}+m_\pi  \approx 1548.6$~MeV with an Euclidean mass parameter $m_\rho = 1408.2$ MeV that is also much larger than the corresponding pole mass.

Because of the necessity to suppress capture processes involving spurious longitudinal vector components at higher temperatures, as relevant in the next subsection, we have reduced $\lambda_\Lambda$ by about an order of magnitude here already, to further increase the longitudinal masses as mentioned in Sec.~\ref{subsec:frg_propagator}. Readjusting the values of the coupling $g$ and UV mass parameter $m_{v,\Lambda} $, by using Set 2 instead of Set 1 here, then restores the physical pole masses of $\rho$ and $a_1$ again, cf.~\Tab{tab:masses}. While this would not be necessary at low temperatures, we avoid changing the UV parameters at higher temperatures in this way.

The curvature masses of pion and sigma meson are the same as in \Eq{eq:masses_values} and so is the quark mass.  The decay of a $\rho$ into pion pair therefore starts at a reasonable threshold of $280.8$ MeV here (and $a_1 \to \sigma + \pi$ at $697.5$~MeV). The decays into quark-antiquark pairs starting at $600$~MeV are not suppressed since there is no confinement in our model (a simple Polyakov-loop enhancement of the model will not do the job). However, all processes give positive contributions and therefore lead to positive (axial-)vector-meson spectral functions.

Putting all contributions for the real and imaginary parts of $\Gamma^{(2),R}$ together gives the vacuum spectral functions of the $\rho$ and $a_1$ meson shown in \Fig{fig:sf_linear}. The peaks are in both cases located at the pole masses where the possible decays lead to rather broad structures in both spectral functions.

\begin{figure}[b]
	\includegraphics[width=0.49\textwidth]{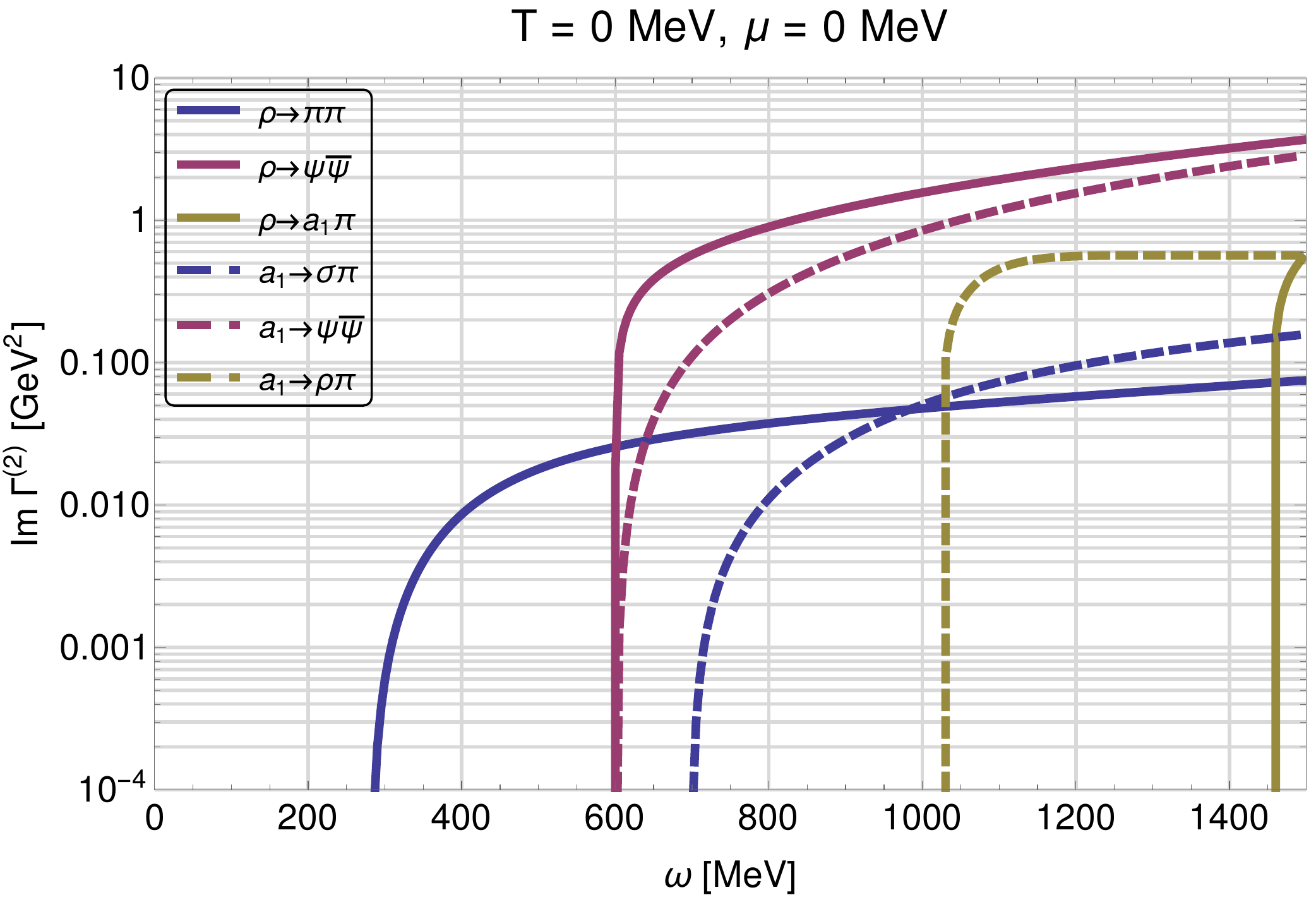}
	\caption{(color online) Imaginary parts of the retarded two-point functions of $\rho$ (solid) and $a_1$ (dashed) over $\omega$ as in Fig.~\ref{fig:imag}, but here evaluated at the scale-dependent $\sigma = \sigma_{0,k} $ using the parameters of Set 3.}\label{fig:imag_rm} 
\end{figure}

A rather simple improvement of the unphysically large two-particle thresholds involving decays into $\rho$ and $a_1$ can be obtained by giving up the unified treatment of (pseudo-)scalar and (axial-)vector mesons in the following way: We use the Euclidean input evaluated at the $k$-dependent minimum $\sigma_{0,k}$ of the scale-dependent effective potential as we did for the flow of the mass parameters in the last subsection, when solving  the aFRG flow equations for the $\rho$ and $a_1$ two-point functions $\Gamma^{(2),R}_{\rho/a_1}(\omega)  $. For the scale-dependent effective potential itself, on the other hand, we stick to the grid technique in the field variable $\phi^2$, to include the order-parameter fluctuations due to collective excitations in the $\sigma$-$\pi$ system as we have been doing so far.

\begin{figure}[t]
	\includegraphics[width=0.49\textwidth]{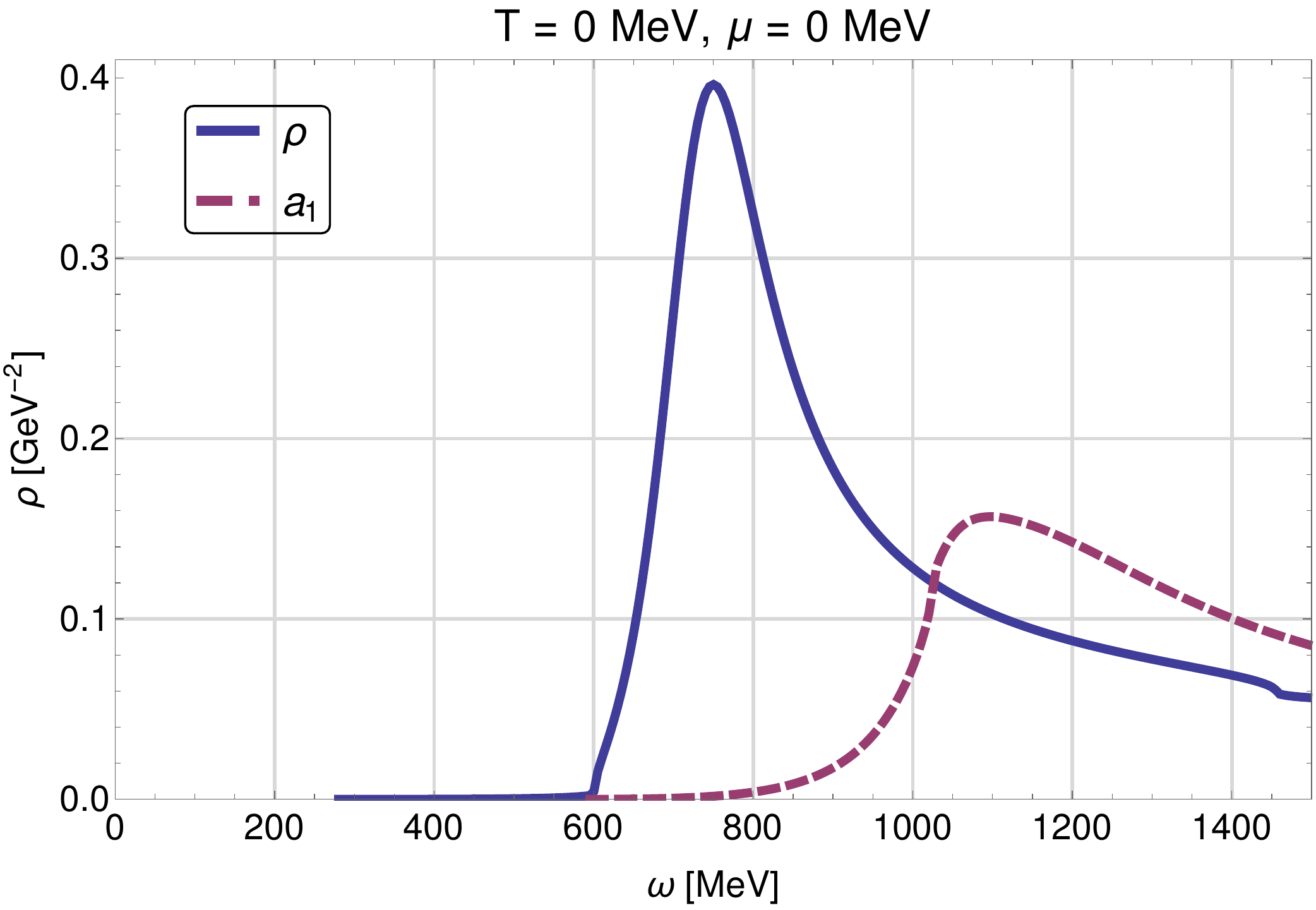}
	\caption{(color online) Vacuum spectral functions of $\rho$ (solid) and $a_1$ (dashed) over $\omega$ as in Fig.~\ref{fig:sf_linear}, but here evaluated at the scale-dependent $\sigma = \sigma_{0,k} $ using the parameters of Set 3.}\label{fig:sf_linear_rm} 
\end{figure}

Because this changes the pole masses of $\rho$ and $a_1$, we have to change their coupling $g$ and UV mass parameter $m_{v,\Lambda}$ once again to compensate this, which is achieved here by using the parameters of Set 3 in Table~\ref{tab:parameters}. The imaginary parts of $\Gamma^{(2),R}_{\rho/a_1}(\omega)$ and the $\rho$ and $a_1$  spectral functions that result from this procedure are shown in Figs.~\ref{fig:imag_rm} and \ref{fig:sf_linear_rm}.

While there are no qualitative changes as compared to the corresponding previous results in Figs.~\ref{fig:imag} and \ref{fig:sf_linear}, with the Euclidean mass parameters 
$m_{\rho}=885.5$~MeV and $m_{a_1}=1316.2 $~MeV, the thresholds of  $a_1 \to \rho + \pi $ and $\rho\to a_1 + \pi$ have now moved down to 
$m_{\rho}+m_\pi \approx 1025.9$~MeV and $m_{a_1}+m_\pi \approx 1456.6$~MeV, respectively. They are still not at the corresponding sums of pole masses, but the differences are considerably reduced by this method of evaluating the Euclidean input for the aFRG calculations of the $\rho$ and $a_1$ correlators at the scale-dependent minimum $\sigma_{0,k}$ of the effective potential. In contrast to the decays involving (axial-)vectors, this modification has no effect on quark-antiquark thresholds or those involving only decays into $\sigma$ and $\pi$, because the relevant (curvature) masses are the same in the infrared.

\subsection{In-medium results}\label{subsec:results_medium}

In order to study the in-medium behavior of the $\rho$ and $a_1$ meson spectral functions across the entire phase diagram of the model, the present setup is straightforwardly extended to finite temperature and quark chemical potential. In the present subsection all calculations are based on the scale-dependent Euclidean input evaluated at the fixed value in field space corresponding to the infrared minimum at $\sigma_0 = 93$~MeV again, and with the parameters of Set 2 in Table \ref{tab:parameters}.    

Because the Euclidean mass parameters determine the thresholds of all processes, we first look at their $T$ and $\mu$-dependencies which are qualitatively the same as in \cite{Jung:2016yxl} where the corresponding phase diagram is shown as well. 

\begin{figure}[t]
	\includegraphics[width=0.49\textwidth]{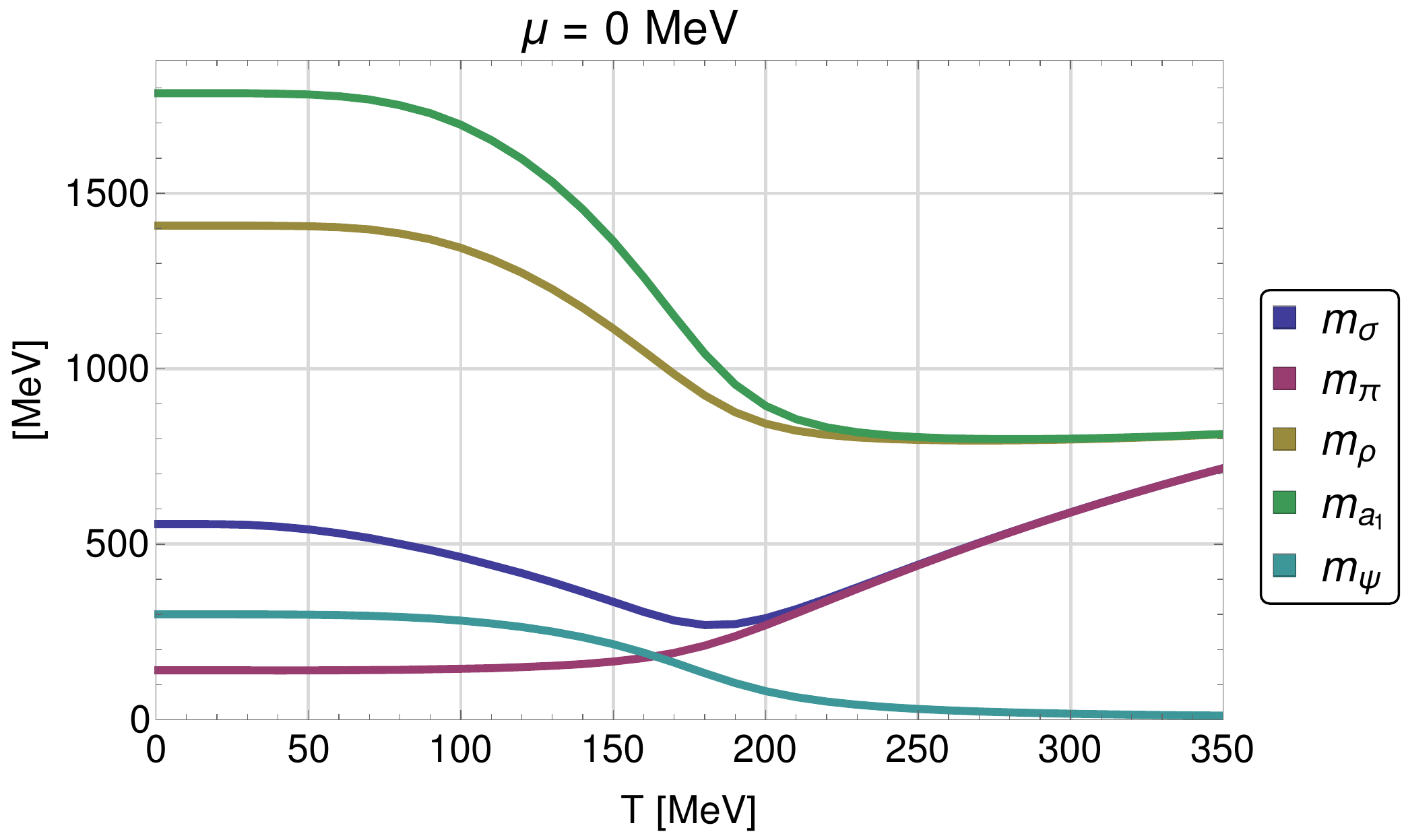}
	\caption{(color online) Euclidean mass parameters over temperature at vanishing chemical potential.}\label{fig:masses_T} 
\end{figure}

In \Fig{fig:masses_T} we show the Euclidean mass parameters as functions of temperature at vanishing chemical potential. The qualitative behavior roughly resembles that of the scale-dependent mass parameters, as obtained by the successive inclusion of quantum fluctuations, at zero temperature in \Fig{fig:masses}.

With spontaneously broken chiral symmetry in the vacuum, the zero-temperature  masses of the chiral partners are split up, and the quarks are massive, here with a constituent mass of $300$~MeV. The $T=0$ values of all mass parameters in \Fig{fig:masses_T} agree with those determining the thresholds in \Fig{fig:imag}. 
Increasing the temperature, chiral symmetry gets gradually restored as indicated by the melting of the order parameter $\sigma_0(T)$ (and so of the quark mass $m_{\psi}=h_s\,\sigma_0$). At temperatures beyond the crossover region the mass parameters of the chiral partners become rapidly degenerate.

\begin{figure}[b]
	\includegraphics[width=0.49\textwidth]{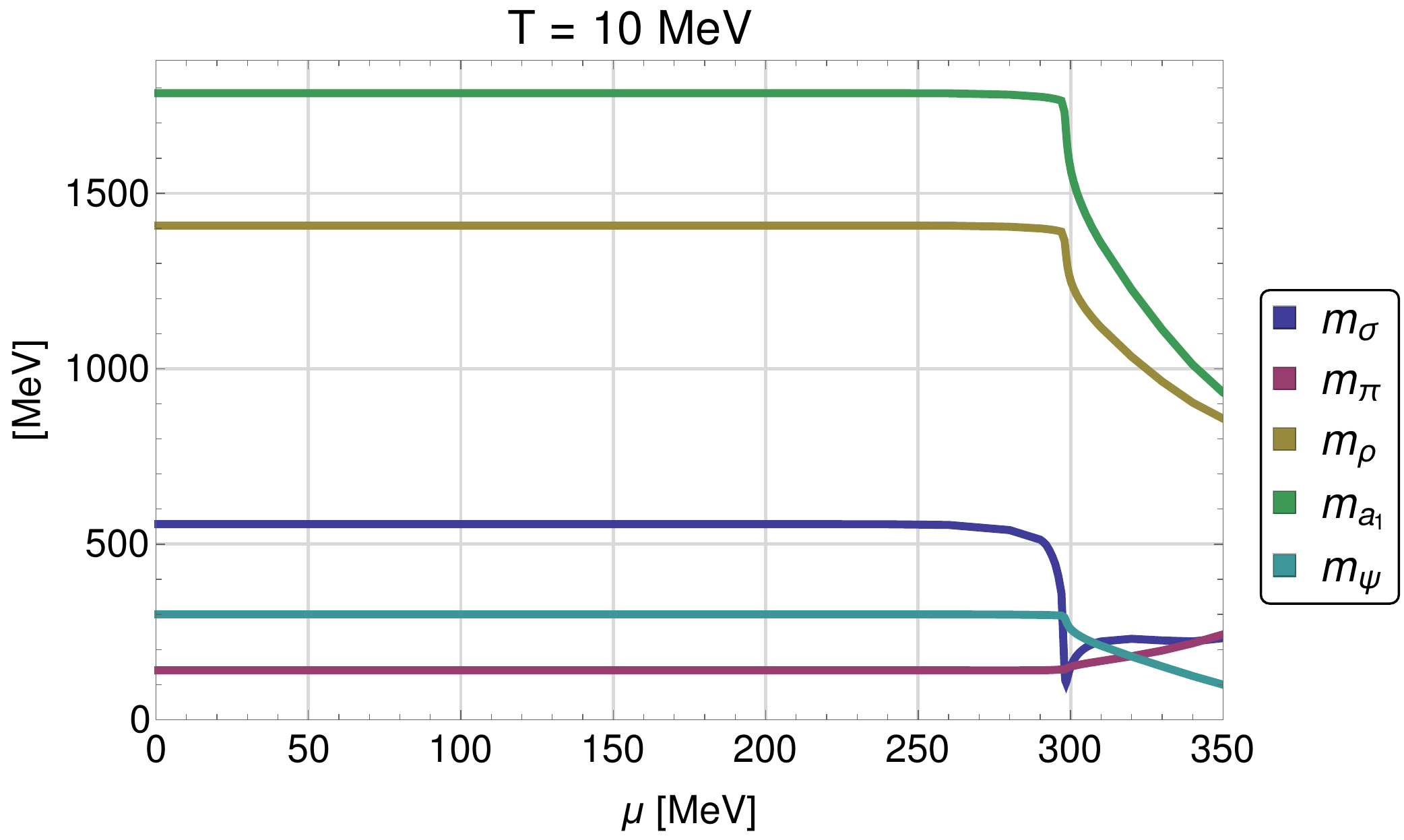}
	\caption{(color online) Euclidean mass parameters over chemical potential at $T=10$~MeV, approximately across the CEP.}\label{fig:masses_mu} 
\end{figure}
 
\begin{figure*}
\begin{center}
	\includegraphics[width=0.45\textwidth]{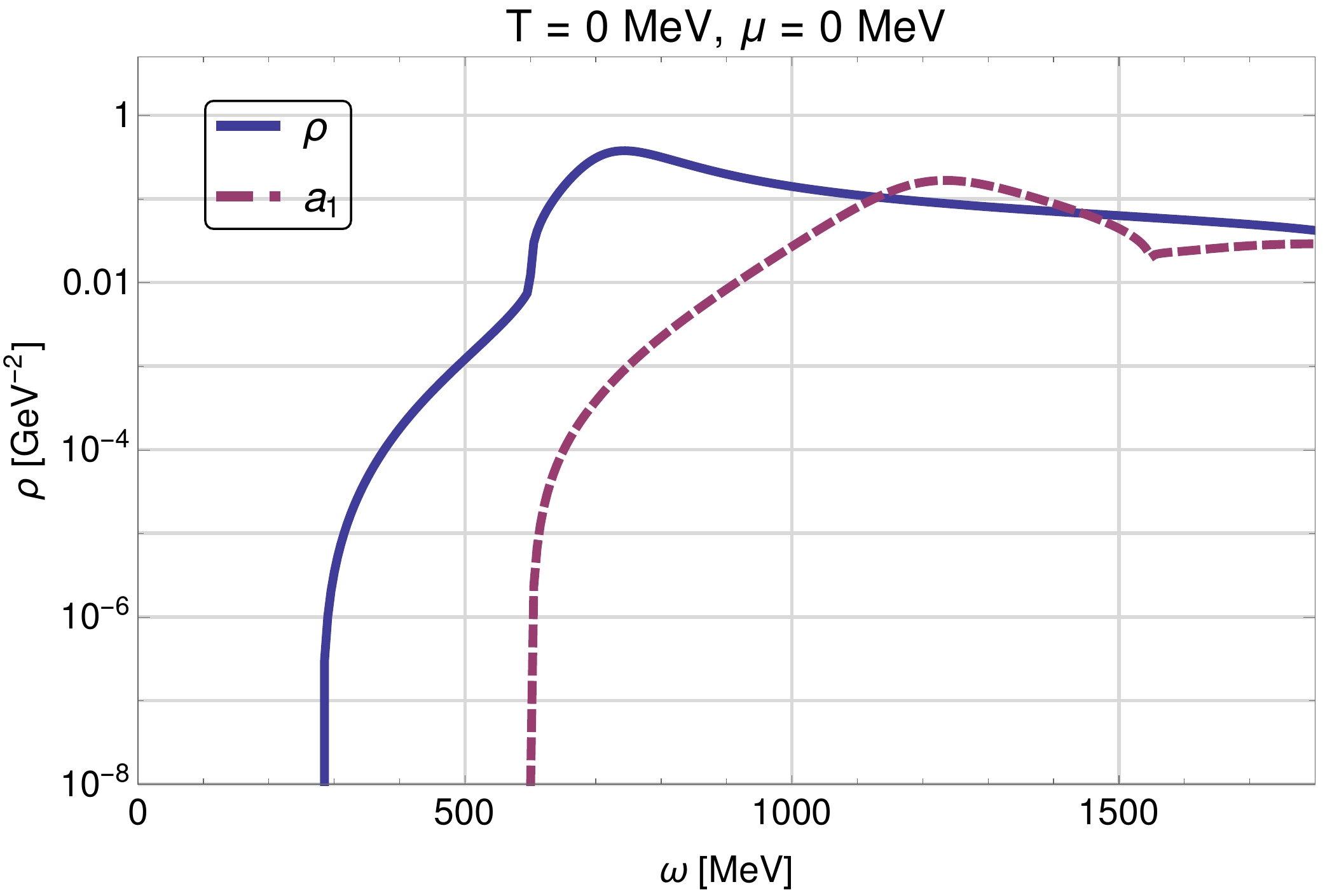}\hspace{.5cm}
	\includegraphics[width=0.45\textwidth]{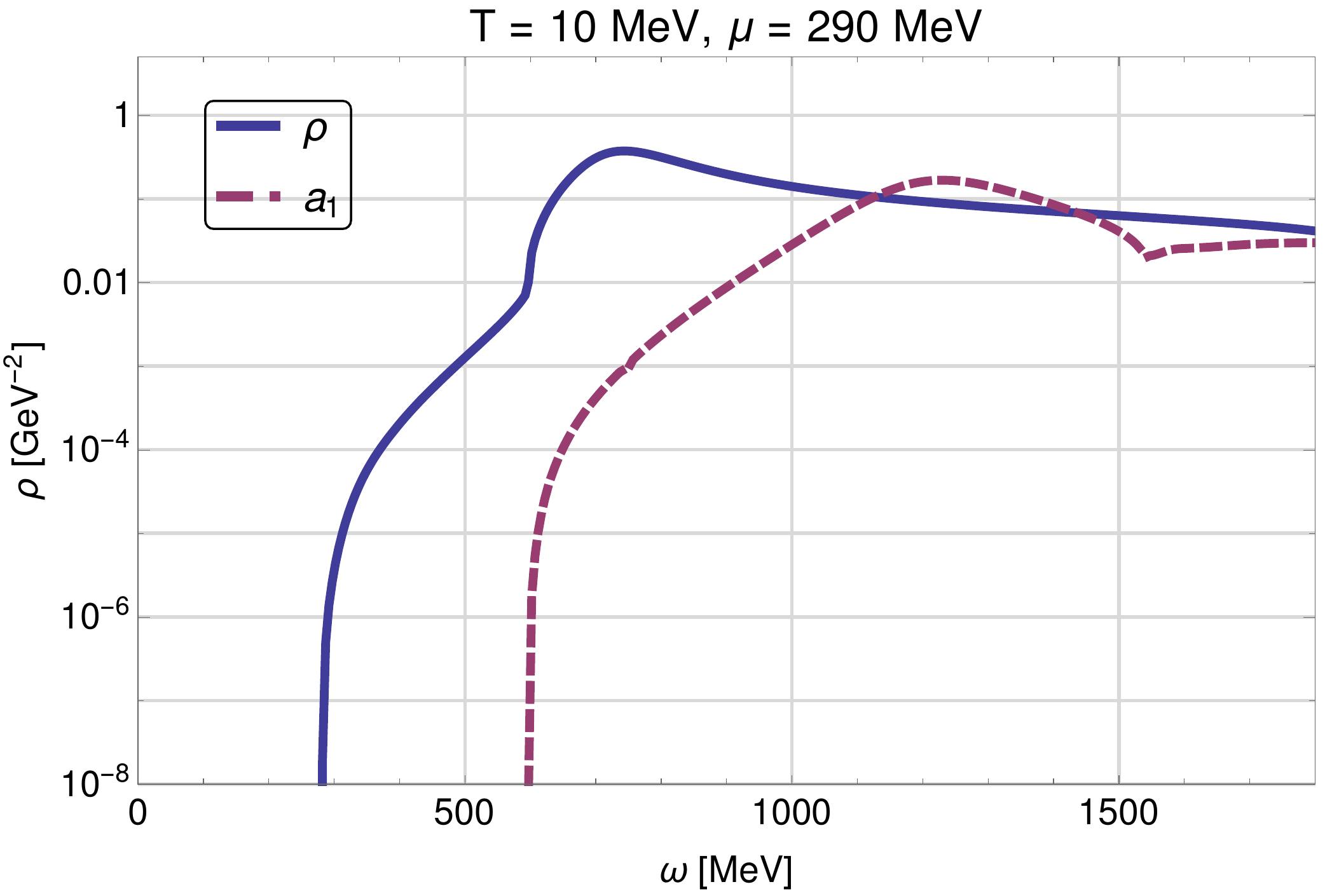}\\\vspace{5mm}
	\includegraphics[width=0.45\textwidth]{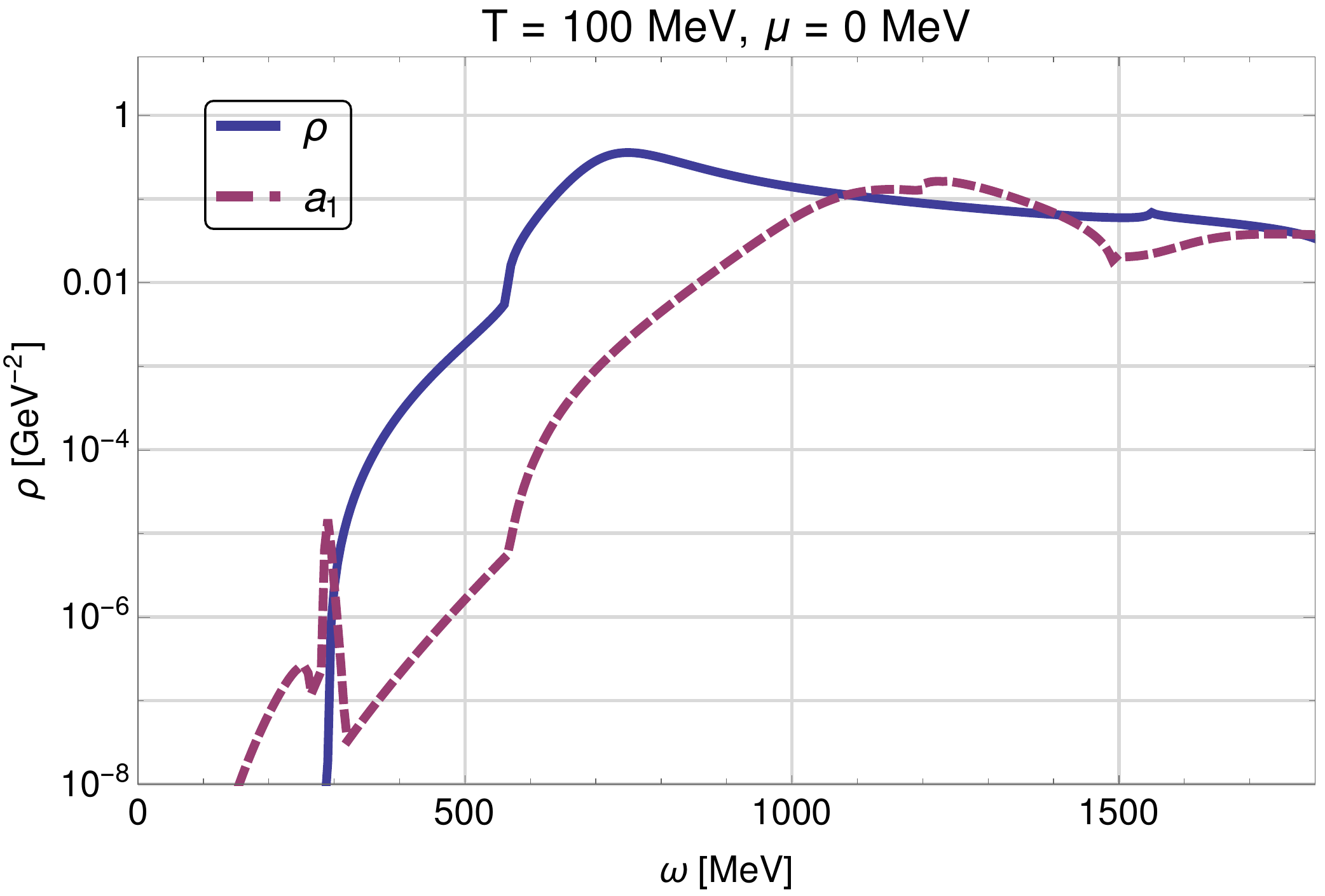}\hspace{.5cm}
	\includegraphics[width=0.45\textwidth]{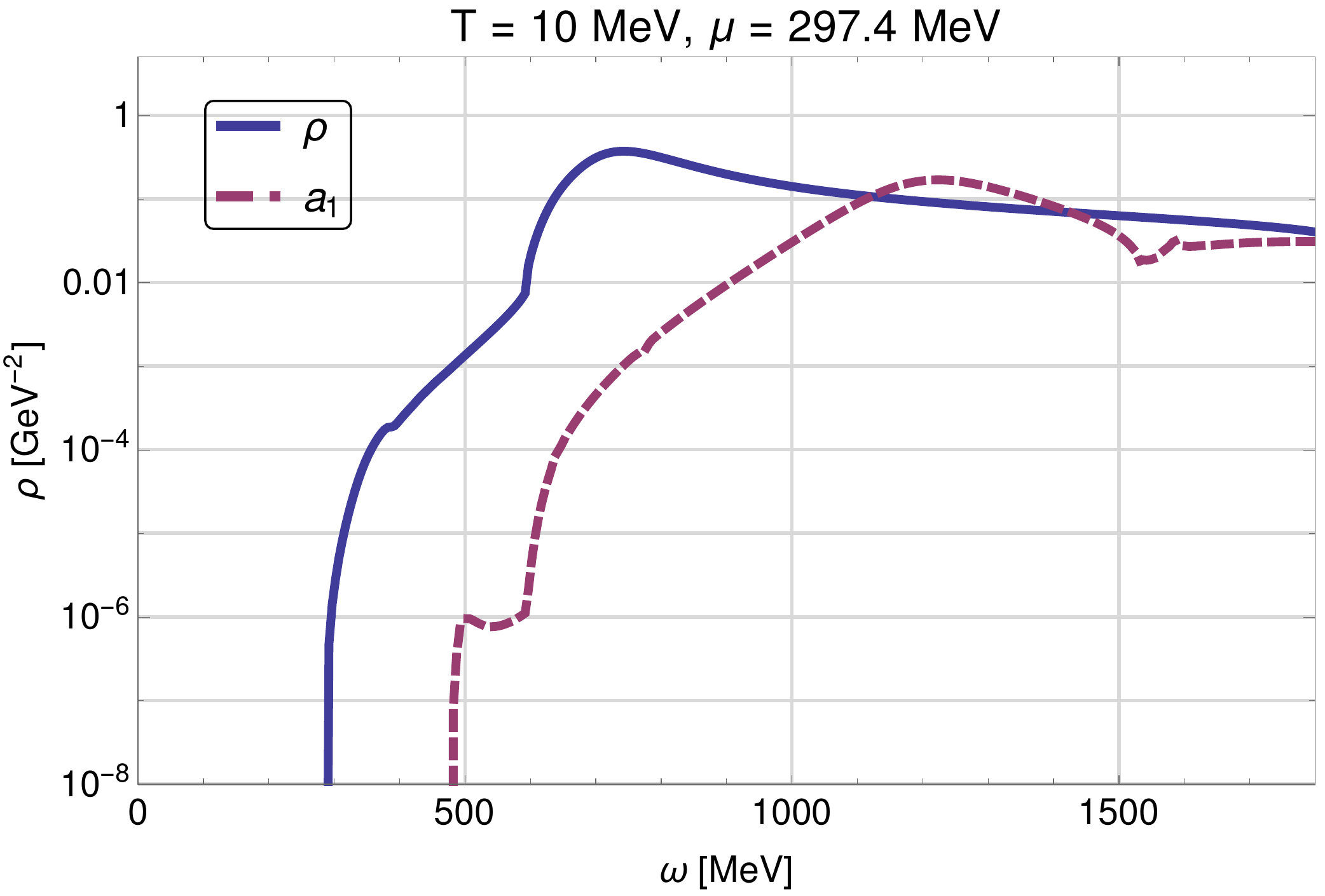}\\\vspace{5mm}
	\includegraphics[width=0.45\textwidth]{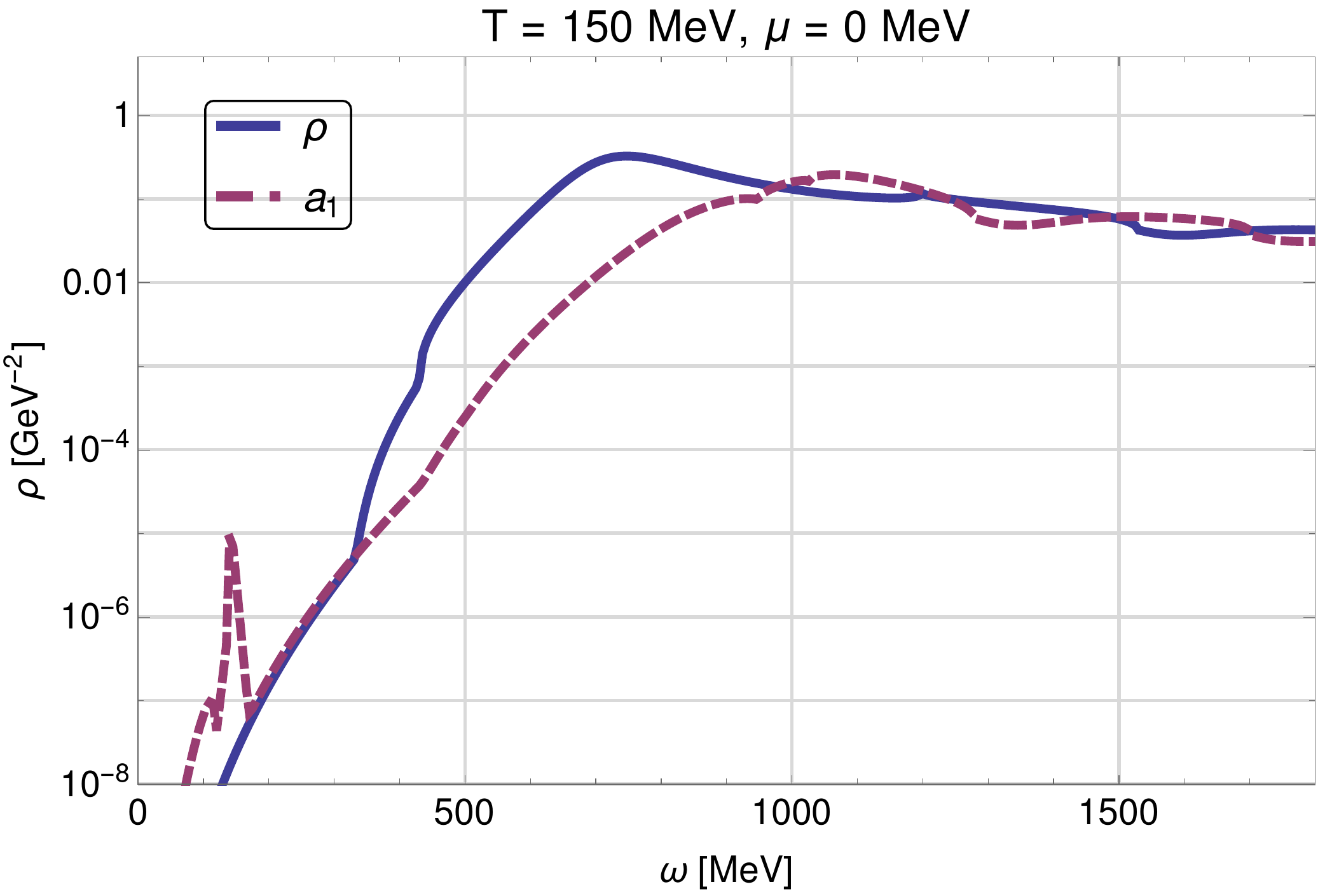}\hspace{.5cm}
	\includegraphics[width=0.45\textwidth]{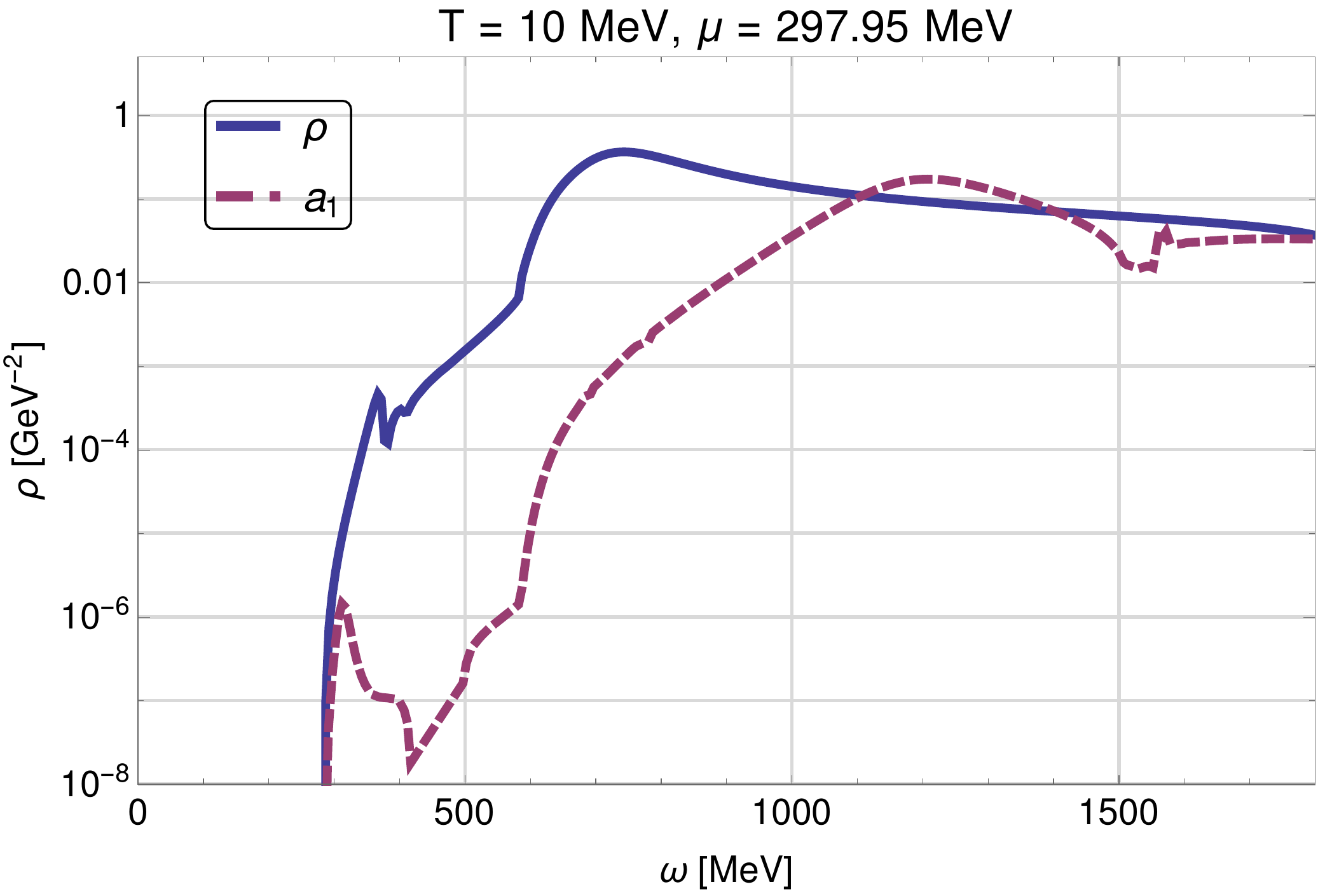}\\\vspace{5mm}
	\includegraphics[width=0.45\textwidth]{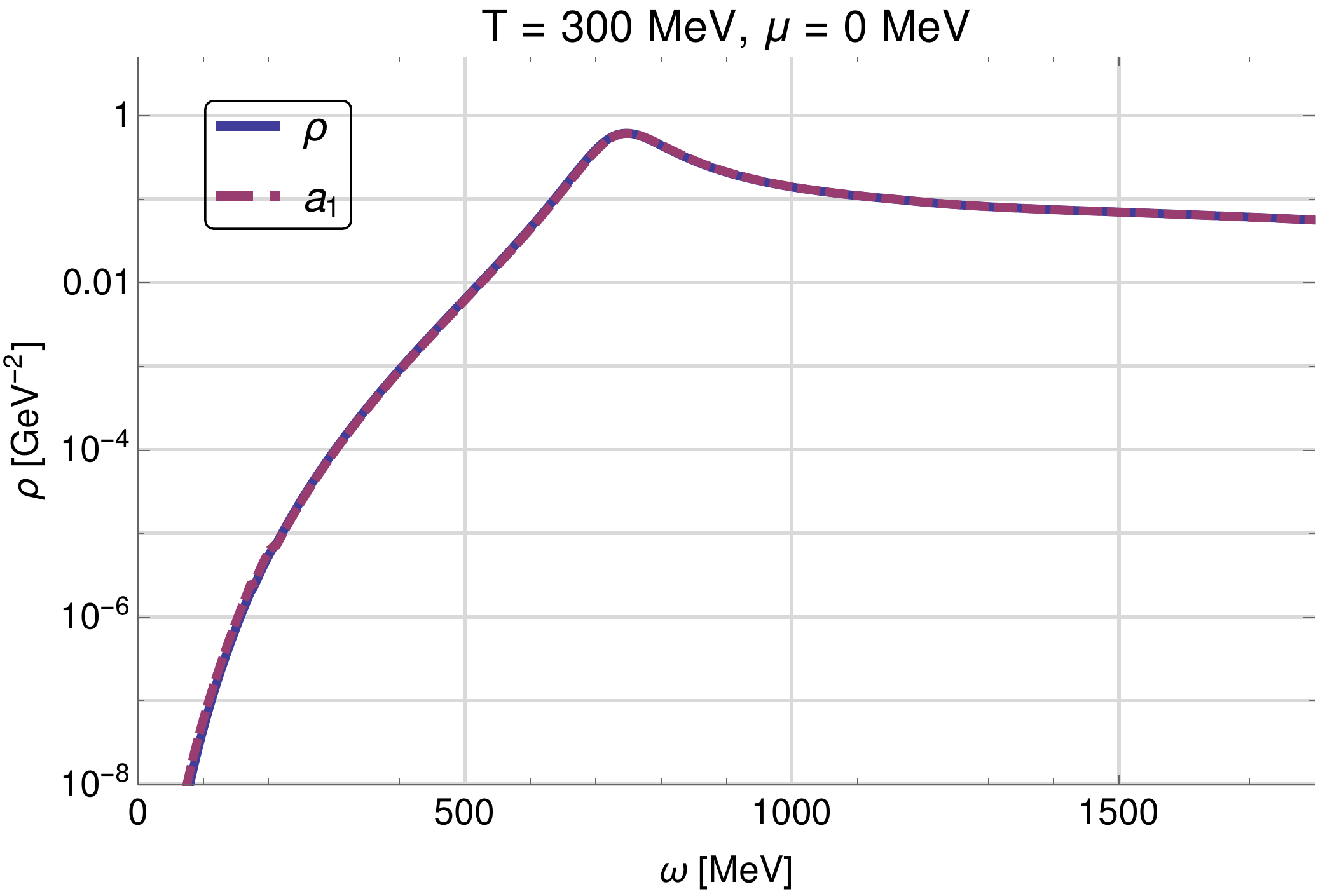}\hspace{.5cm}
	\includegraphics[width=0.45\textwidth]{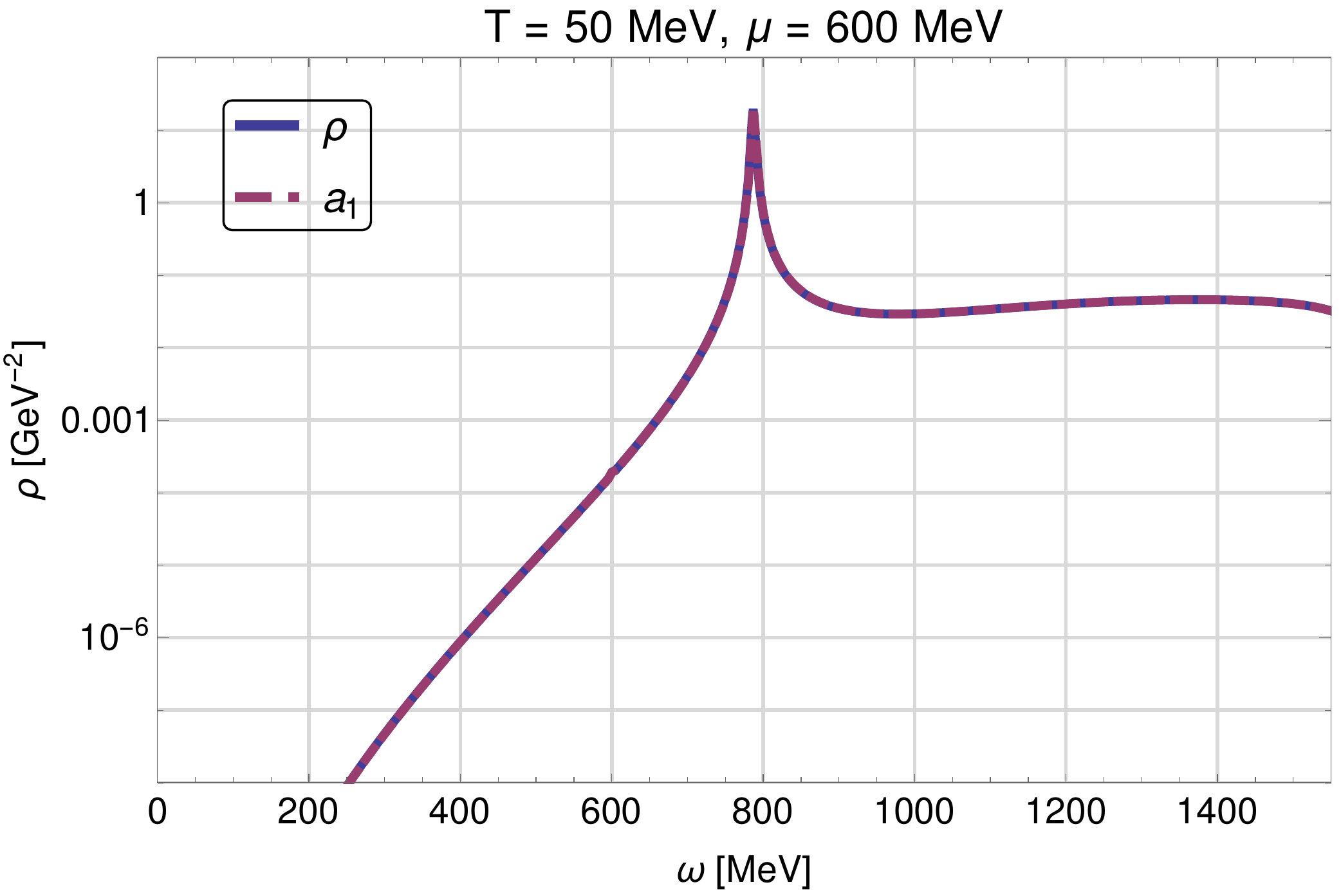}
	\caption{(color online) In-medium spectral functions of the $\rho$ (solid blue) and $a_1$ meson (dashed red) as a function of external frequency $\omega$ for increasing temperature at $\mu=0$ MeV (left column) and for increasing chemical potential close to the CEP at $T=10$ MeV (right column) plotted in logarithmic scales. }\label{fig:spectral_functions_temp_mu}
\end{center}
\end{figure*}

The chemical-potential dependence of these Euclidean masses, at a fixed temperature of $T=10$~MeV which is approximately that of the CEP with our model parameters, is shown in \Fig{fig:masses_mu}. We observe that they remain constant over a wide range of $\mu$, which might seem curious at first, just as the dog that did nothing in the night-time in {\em The Adventure of Silver Blaze} \cite{Cohen2003}, and which is due the temperature being so low here. At $\mu\approx 298$~MeV, however, all masses except that of the pion drop abruptly. The sigma becomes almost massless indicating the proximity of the critical endpoint which represents a second-order phase transition with $Z_2$-universality where only the correlation length of the $\sigma$-field diverges. For very large $\mu$ we again see the degeneracy of the masses of the chiral partners now reflecting the gradual restoration of chiral symmetry with density inside the high-density phase (here of self-bound quark matter).

We are now ready to discuss the temperature and chemical-potential dependence of the  spectral functions of $\rho$ and $a_1$ mesons shown in \Fig{fig:spectral_functions_temp_mu}. In the present setup, the following time-like decay and capture processes are possible for an off-shell $\rho^*$ vector meson,
\begin{alignat}{2}
&\rho^* \rightarrow \psi+\bar{\psi}\,,\quad &&\rho^* \rightarrow \pi+\pi\,,\nonumber\\
  &\rho^* \rightarrow a_1+\pi\,,\quad &&\rho^*+\pi \rightarrow a_1\,,
\end{alignat}
and for an off-shell $a_1^*$ axial-vector meson,
\begin{alignat}{2}
&a_1^*\rightarrow \psi+\bar{\psi}\,,\quad &&a_1^*+\sigma\rightarrow \pi\,,\nonumber\\
&a_1^*\rightarrow \pi+\sigma\,,\quad &&a_1^*+\pi\rightarrow \sigma\,,\nonumber\\
  &a_1^*\rightarrow \rho+\pi\,,\quad &&a_1^*+\pi\rightarrow \rho\,,
  \nonumber\\
  &a_1^*\rightarrow a_1+\sigma\,,\quad &&a_1^*+\sigma\rightarrow a_1\,,
\end{alignat}
where each particular process is possible only if it is energetically allowed, i.e.~a decay process only for energies  $\omega$ in the off-shell meson correlator with  \mbox{$\omega\geq m_{\alpha}+m_{\beta}$}, and a capture process for \mbox{$\omega+m_{\alpha}\leq m_{\beta}$} which requires  $m_\beta > m_\alpha$ in addition. We have therefore not listed here the capture processes $\rho^*+a_1 \rightarrow \pi $, $a_1^*+\rho\rightarrow \pi$ and $a_1^*+a_1\rightarrow \sigma $ where this never occurs, while it does happen that $m_\sigma < m_\pi$ very close to the critical endpoint as seen in Fig.~\ref{fig:masses_mu}. 

\begin{figure}[b]
	\includegraphics[width=0.49\textwidth]{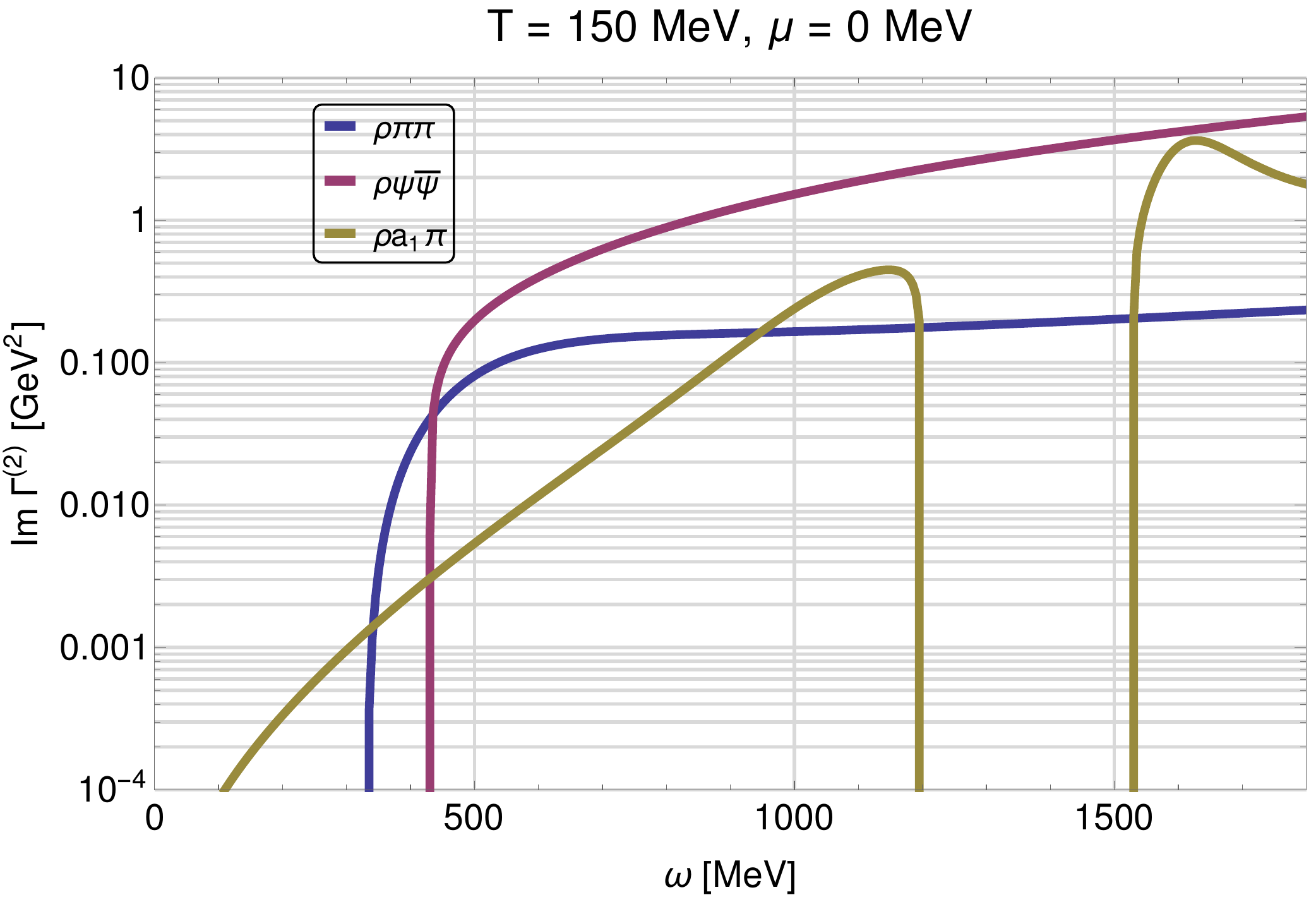}
	\caption{(color online) Imaginary parts of the retarded two-point function of the $\rho$ meson for every process separately as a function of external frequency $\omega$ at $T=150$~MeV (evaluated at fixed IR minimum $\sigma = \sigma_0 $ with parameters of Set 2).}\label{fig:imag_rho_T} 
\end{figure}

In the vacuum only the decay processes contribute, cf.~\Fig{fig:imag}, and thus the starting thresholds are given by the decays into the light (pseudo-)scalar mesons in both spectral functions. Here, essentially the only new feature as compared to the previous study in \cite{Jung:2016yxl} is the decay  $a_1 \to \rho + \pi $ as discussed in the previous subsection. Since the decays into quark-antiquark pairs yield rather large contributions, cf.~Ref.~\cite{Jung:2016yxl}, both spectral function always broaden in all $T$ and $\mu$ cases for $\omega\geq2 m_\psi$. However, with increasing temperature (from top to bottom in the left column of \Fig{fig:spectral_functions_temp_mu}) we see that the various capture processes start to contribute, giving rise to an increase in both spectral functions especially at low external frequencies, where in the $a_1$ spectral function the van Hove peak appears that was already observed in \cite{Jung:2016yxl}. At $T=300$ MeV the masses of the chiral partners are degenerate and the quarks become the lightest degrees of freedom leading to broad and fully degenerate spectral functions of $\rho$ and $a_1$ mesons as expected in the chirally restored phase. 

\begin{figure}[b]
	\includegraphics[width=0.49\textwidth]{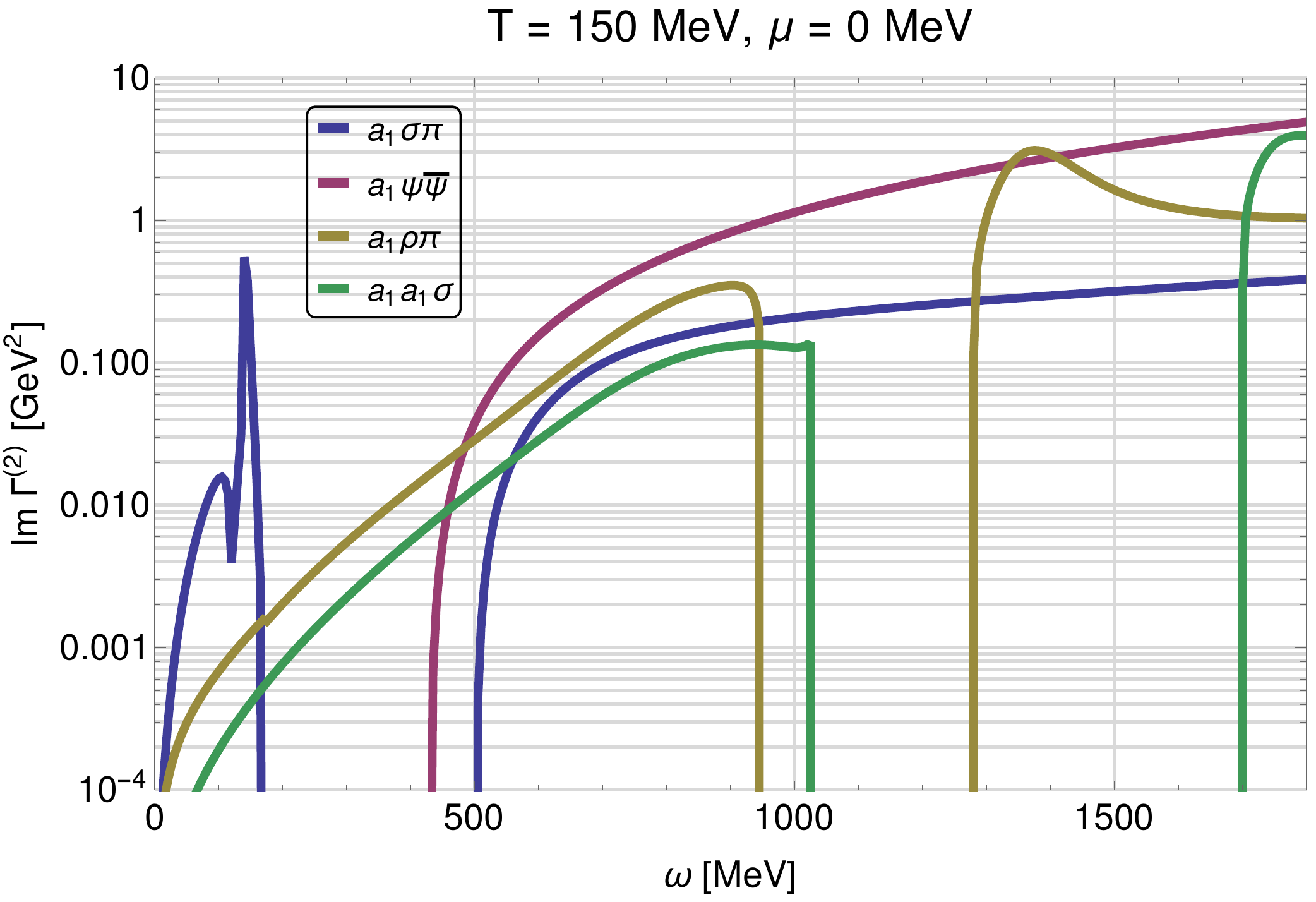}
	\caption{(color online) Imaginary parts of the retarded two-point function of the $a_1$ meson for every process separately as a function of external frequency $\omega$ at $T=150$~MeV (evaluated at fixed IR minimum $\sigma = \sigma_0 $ with parameters of Set 2).}\label{fig:imag_a1_T} 
\end{figure}

The most noticeable differences with respect to the previous results in \cite{Jung:2016yxl} are the new capture processes involving vector mesons, i.e.~$\rho^* + \pi \to a_1$, $a_1^* + \pi \to \rho$ and $a_1^* + \sigma \to a_1$ which fill up the gap between capture and decay processes when only the light (pseudo-)scalars are involved. This is demonstrated in Figs.~\ref{fig:imag_rho_T} and \ref{fig:imag_a1_T} where we plot the individual imaginary parts of the $\rho$ and the $a_1$ two-point function separately that altogether contribute to their spectral functions at $T=150$~MeV as an example. 

Turning on the chemical potential (as done in the right column of \Fig{fig:spectral_functions_temp_mu}), when the low-temperature system at $T=10$~MeV eventually starts to react, beyond $\mu \simeq 290$~MeV, we observe modifications in the $a_1$-meson spectral function close to the critical endpoint located at \mbox{$\mu_{\mathrm{CEP}}\approx 298$ MeV}. While the $\rho$ meson spectral function remains qualitatively unchanged, the dropping threshold for \mbox{$a_1^*\rightarrow \pi + \sigma$} and the related peak in the $a_1$ spectral function are due to the dropping sigma mass in this critical region and can thus serve as a signature for the CEP. As compared to the previous study in \cite{Jung:2016yxl} this signal got somewhat washed out by the additionally possible processes, unfortunately, but the fact that it is robust enough to still be visible seems at least encouraging for further studies.   At large $\mu$ we again observe the full degeneracy of both spectral functions with gradual chiral symmetry restoration inside the high density phase.

\begin{figure}[t]
	\includegraphics[width=0.49\textwidth]{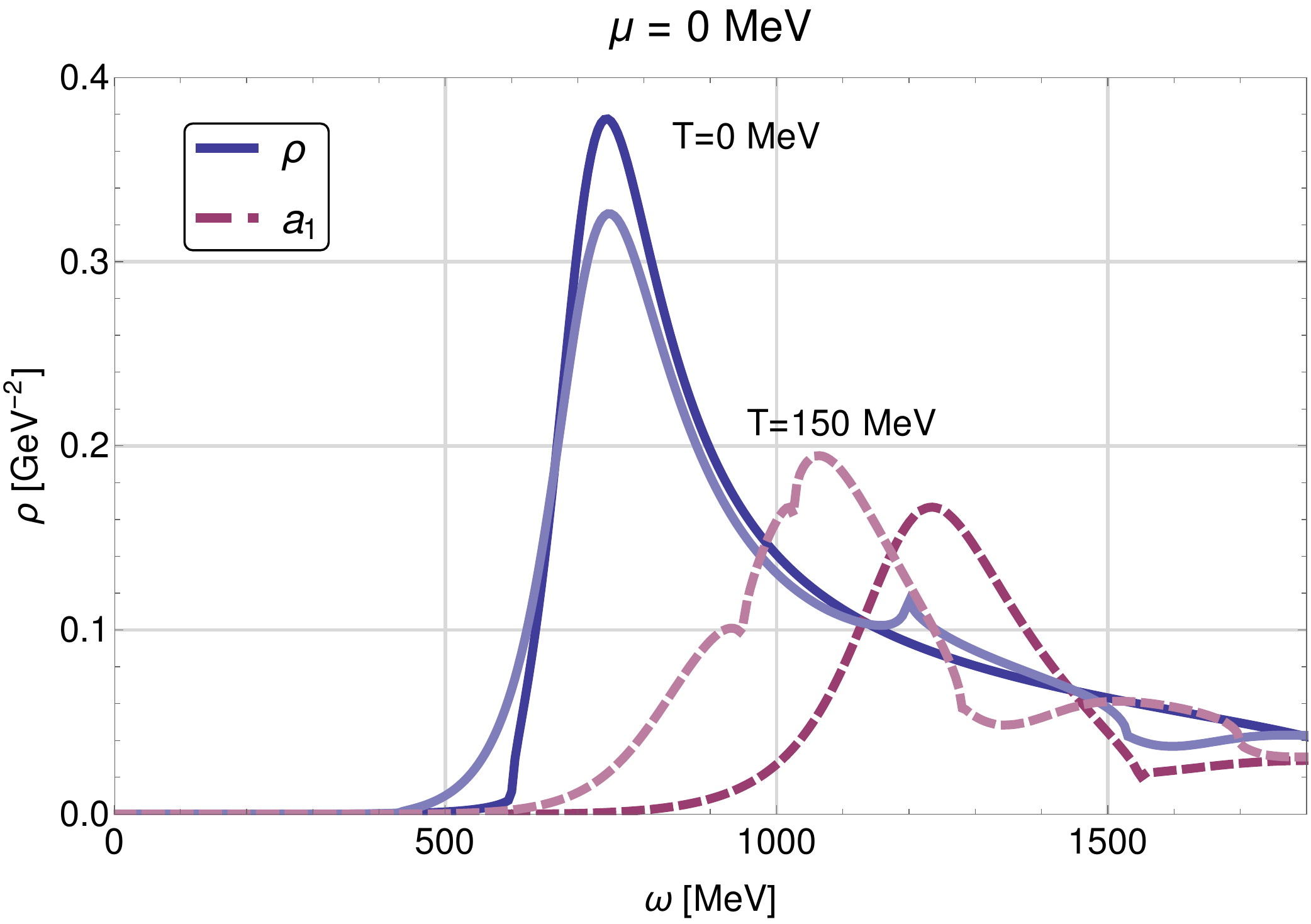}
	\caption{(color online) Spectral functions of the $\rho$ (solid blue) and $a_1$ meson (dashed red) as a function of external frequency $\omega$ for $T=0$ MeV (dark) and $T=150$ MeV (light).}\label{fig:spectral_functions_twice} 
\end{figure}

\begin{figure}[b]
	\includegraphics[width=0.49\textwidth]{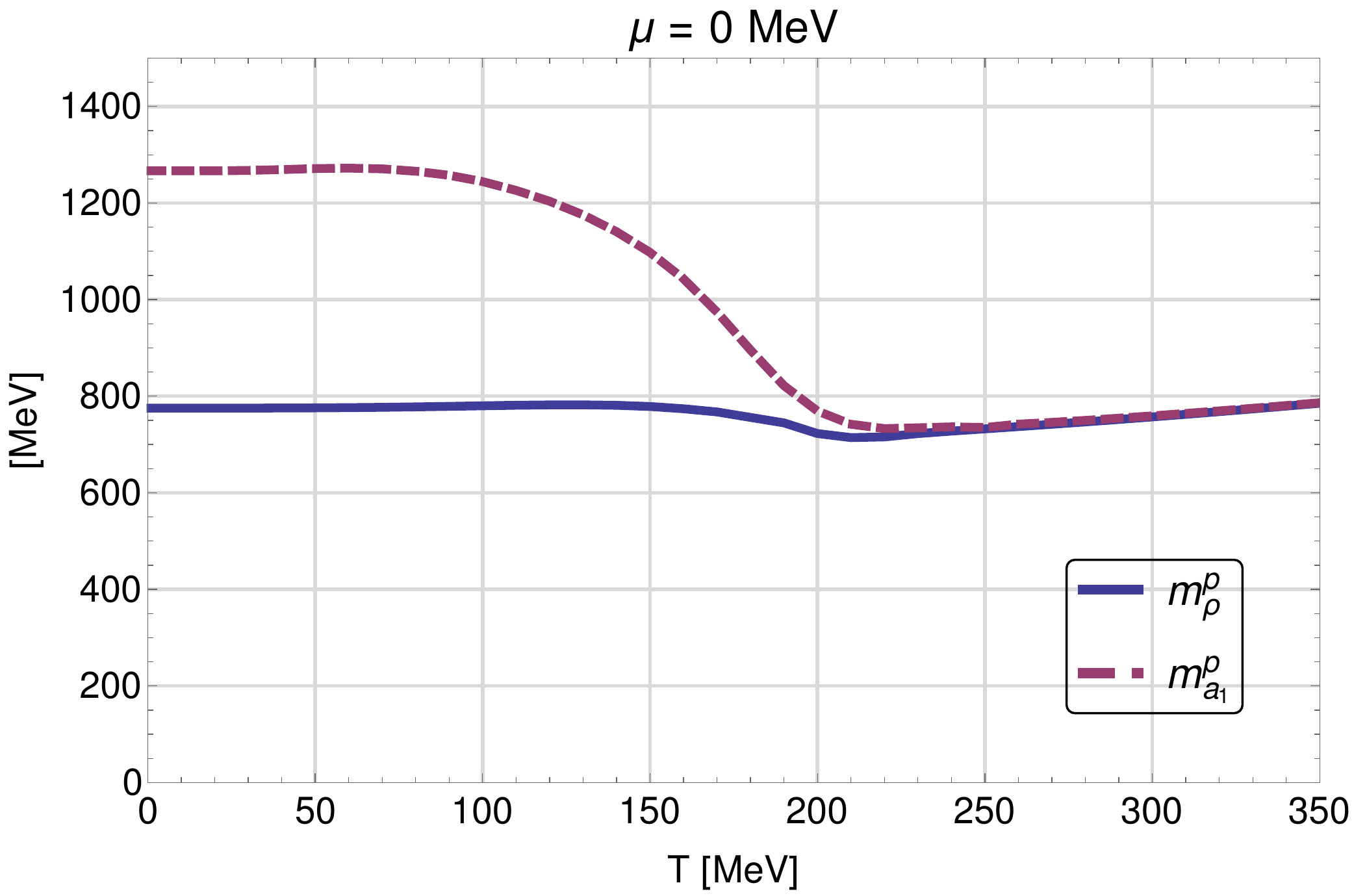}
	\caption{(color online) Temperature dependence of the pole masses of $\rho$ (solid blue) and $a_1$ (dashed red) meson at $\mu = 0$.}\label{fig:pole_masses_t} 
\end{figure}

To summarize, the modifications in the thermal medium are once more exemplified in \Fig{fig:spectral_functions_twice} where the spectral functions at \mbox{$T=0$ MeV} and \mbox{$T=150$ MeV} are compared to one another in a linear plot. The increasing temperature leads to a decreasing $a_1$ pole mass, whereas the peak of the $\rho$ meson stays put as it melts down with temperature. This effect continues with further increasing the temperature until both spectral functions are fully degenerate. 

As in the previous study of Ref.~\cite{Jung:2016yxl} this is once again in line with the melting-$\rho$ scenario \cite{Rapp:1999ej,Rapp:2009yu,Hohler:2015iba,Rennecke:2015eba}. That the position of the $\rho$-meson resonance is even less temperature dependent here than in the previous study in \cite{Jung:2016yxl} is seen in Fig.~\ref{fig:pole_masses_t} where we plot our estimates of the real parts of the positions of the corresponding poles on the unphysical Riemann sheet as the physical pole masses of $\rho$ and $a_1$ mesons over the chiral crossover with increasing temperature.

\begin{figure}[t!]
	\includegraphics[width=0.49\textwidth]{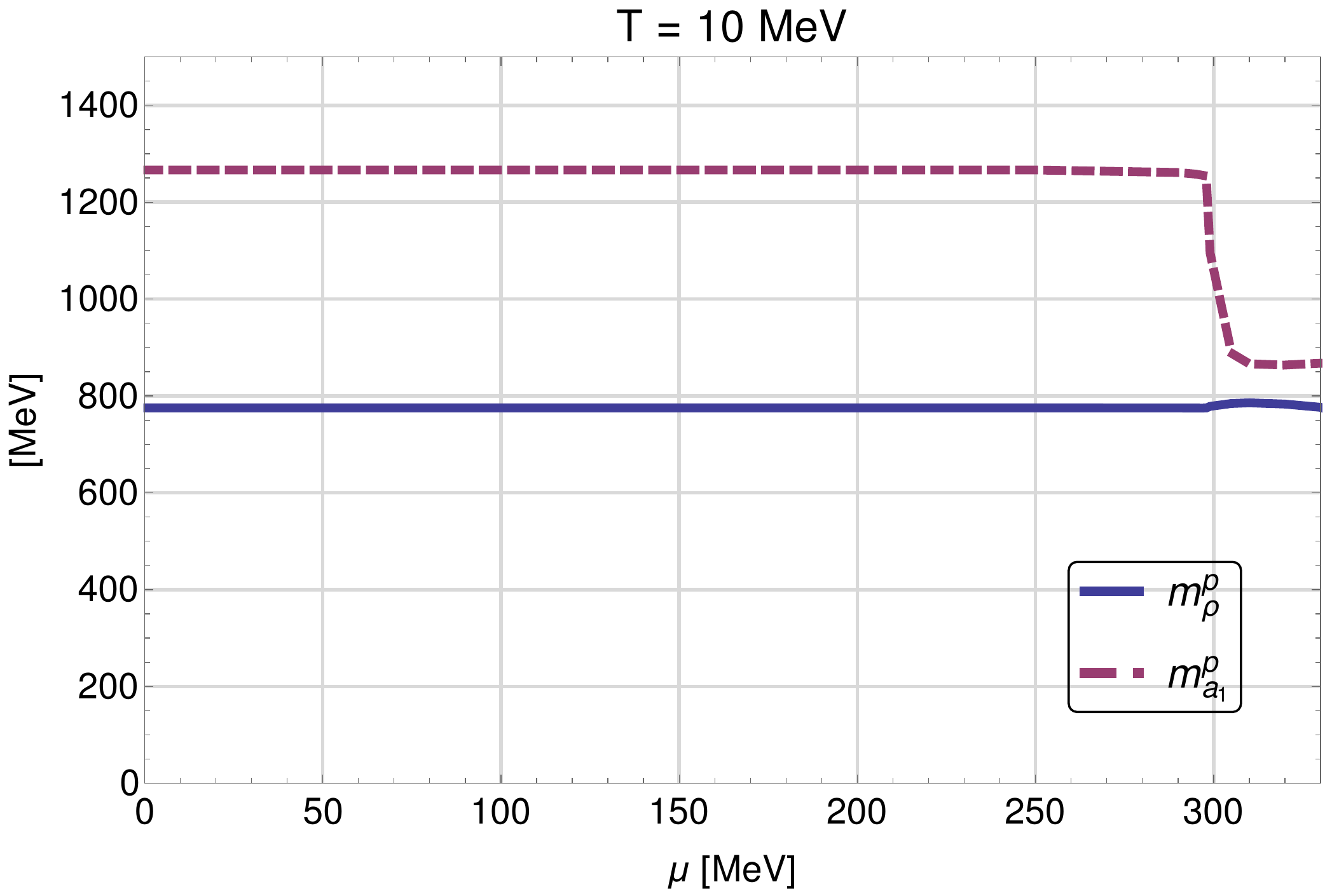}
	\caption{(color online) Chemical potential dependence of the pole masses of $\rho$ (solid blue) and $a_1$ (dashed red) meson at $T = 10$~MeV.}\label{fig:pole_masses_mu} 
\end{figure}

The corresponding dependence of the same physical pole masses of $\rho$ and $a_1$ on the chemical potential at the constant temperature of $T= 10$~MeV across the critical endpoint is shown in Fig.~\ref{fig:pole_masses_mu}. The pole mass of the $a_1$ also drops considerably in the critical region whereas that of the $\rho$ meson hardly shows much change here either.

\section{Summary and Conclusion}\label{sec:summary}

In this paper we have presented our results from computing in-medium vector and axial-vector meson spectral functions within an effective chirally gauged low-energy theory. Extending the previous study of Ref.~\cite{Jung:2016yxl} we have thereby for the first time also included the fluctuations due to these vector and axial-vector mesons themselves in this non-perturbative computational framework based on the functional renormalization group (FRG). More specifically, the originally Euclidean FRG flow equations for two-point correlation functions at finite temperature and density are first analytically continued, before they are solved to directly compute retarded Green functions within this {\em analytically continued} aFRG framework.

In order to include the (axial-)vector meson fluctuations in our aFRG flows it was first necessary to understand, how off-shell fluctuations due to these massive (axial-)vectors are described correctly by elementary fields in an effective theory. Unphysical positivity-violating contributions from longitudinal fluctuations need to be suppressed, and spurious massless single-particle contributions to the transverse fluctuations avoided. Our description is based on the known spectral representations for commutators and causal Green functions of the corresponding conserved currents, including seagull and Schwinger terms, which are then translated to the (axial-)vector fields of the effective theory via the current-field identities of vector-meson dominance models. This uniquely fixes the form of the transverse single-particle contributions from the massive (axial-)vector fields inside the loops on the right hand side of the aFRG flow equations. We have implemented these transverse forms, which most importantly do not contain any massless single-particle contributions, together with matching longitudinal components to the (axial-)vector two-point functions which are constructed in a way such that they turn themselves off in the regularized propagators during the flow, as one approaches the limit $k\rightarrow 0$ in the infrared. The procedure of dealing with the longitudinal fluctuations is inspired by the modified Ward identities of FRG flows with current conservation being recovered in the infrared.

Within an effective theory based on the gauged linear sigma model with quarks, we have solved the Euclidean FRG flow for the effective average action in our truncation, as a first step, which is used to determine the thermodynamic potential and the other input needed for the aFRG flows in the second step. For the (axial-)vector fluctuations this input includes scale-dependent mass parameters $m_{v,k}$ and residues $Z_k \equiv m_{v,k}^2/m_{0,k}^2$, i.e.~the strengths of the single-particle contributions, in the spirit of an LPA' truncation. We have explicitly verified the expected ordering $m_{v,k}^2 \leq m_{0,k}^2$ of these mass parameters together with the chiral symmetry breaking pattern in the hierarchy of all Euclidean mass parameters emerging at a typical scale of $k \sim 600$~MeV during the flow.

For comparison, we have computed the spectral functions of $\rho$ and $a_1$ from the corresponding aFRG flows for the same combinations of temperature and chemical potential as in Ref.~\cite{Jung:2016yxl}. While we observe a similar qualitative overall behavior, with the (axial-)vector fluctuations included, there are now additional processes possible which involve the $\rho$ and $a_1$ mesons inside the loops. We have disentangled these various contributions to the spectral functions: For increasing temperature we observe various new capture processes such as $\rho^*+\pi \rightarrow a_1$, $a_1^*+\pi\rightarrow \rho$ and $a_1^*+\sigma\rightarrow a_1$ to dominate the regime between capture and decay processes involving only (pseudo-)scalars. The corresponding new decay channels on the other hand produce additional, although subdominant structure in the high frequency domain. As before, at about \mbox{$T\simeq 300$ MeV} both spectral functions become completely degenerate.

As a function of chemical potential, the previously observed peak in the critical region, of the $a_1$ spectral function at the dropping  $a_1^*\rightarrow\sigma +\pi$ threshold, is still visible on top of all the additional new structure. Perhaps most importantly for further studies, this thus appears to be a robust signature of the critical endpoint.    

In summary, we have developed and tested a successful effective description of (axial-)vector fluctuations due to off-shell single-particle contributions from $\rho$ and $a_1$ mesons in agreement with the general spectral properties of the corresponding conserved (axial-)vector currents. From a phenomenological point of view, an important next step will be to implement not only such single-particle contributions in the aFRG flows, but an enhanced Ansatz which reflects the full non-trivial spectral properties of a given correlation function, or even a completely self-consistent solution where the non-trivial spectral functions are fed back into the aFRG flows, e.g.~via suitable spectral representations. In our present effective theory description, for example, this should lead to the specific peak of the $a_1$ spectral function in the critical region to become observable also in the $\rho$-meson spectral function. From there on, this signature can then mix into the electromagnetic spectral function as described in Ref.~\cite{Tripolt:2018jre}, and hence eventually become an observable feature of thermal dilepton spectra.

\acknowledgments 
The authors gratefully acknowledge fruitful discussions with Naoto Tanji, Ralf-Arno Tripolt and Jochen Wambach. This work was supported by the Helmholtz International Center (HIC) for FAIR within the LOEWE initiative of the State of Hesse, by the Deutsche For\-schungs\-gemeinschaft (DFG) through the grant CRC-TR 211 ``Strong-interaction matter under extreme conditions,'' and by the German Federal Ministry of Education and Research (BMBF), grant no.~05P18RGFCA.



\appendix

\section{Massive spin-1 particles from anti-symmetric rank-2 tensor fields}\label{AppHodge}

Gasser and Leutwyler have proposed to describe the $\rho$-meson in terms of an anti-symmetric rank-2 tensor field $\rho_{\mu\nu}$ when constructing an effective Lagrangian \cite{Gasser:1983yg} starting with a free kinetic Lagrangian plus mass term of the form
\begin{equation}
  \label{FSL}
  \mathcal L_0^\rho = -\frac{1}{2} (\partial_\mu \rho_{\mu\nu} ) \partial_\sigma \rho_{\sigma\nu}   + \frac{1}{4} \, m_v^2 \,\rho_{\mu\nu} \rho_{\mu\nu} \,. 
\end{equation}
We are considering just a single flavor component without gauging for simplicity. All we want to point out here is that this Lagrangian, when the components of $\rho_{\mu\nu}$ are re-expressed in terms of a conserved 4-vector field, leads to a transverse tree-level two-point function for this vector field in momentum space of the form
\begin{equation}
  \Gamma_{\mu\nu}^{(2)T}(p) = \, -\frac{m_v^2}{p^4} (p^2+m_{v}^2)\,
  \big(p^2 \delta_{\mu\nu} - p_\mu p_\nu\big) \, , \label{G2Vects}
\end{equation}
corresponding to the transverse propagator in Eq.~(\ref{eq:eucl_zuber}) or the corresponding Feynman propagator in (\ref{vectorprop}).

The essential step to see how this form arises is based on Hodge decomposition and inversion of the de Rahm-Laplace operator. We therefore sketch it here using the language of de Rahm cohomology for brevity. We thus describe classically conserved left and right-handed $J^\pm$ or (axial-)vector $J_{V/A}$ currents in terms of one-forms with
\begin{equation}
  J^\pm = \frac{1}{2} (J_V \pm J_A) \,, \;\; \mbox{and} \;\; \delta J^\pm = \delta J_{V/A} = 0\,.
\end{equation}
We do not include anomalies in the discussion. By the Poincar\'e Lemma on convex regions of spacetime (we use the Euclidean signature here) these are then co-exact, i.e.~expressed as exterior co-derivatives of (anti-)selfdual two-forms $F^{\pm}$, with $* F^\pm = \pm F^\pm$, 
\begin{equation}
  \label{coexact}
  J^\pm = \delta F^\pm \,.
\end{equation}
The selfdual and anti-selfdual anti-symmetric rank-2 tensors transform in the $(1,0)$ and $(0,1)$ representations of the Euclidean $O(4)$ (or also the proper orthochronous Lorentz group with $(1,0) \leftrightarrow (0,1)$ under parity, so that
 $(1,0) \oplus (0,1)$ is used for massive spin-1 particles).    

Upon exterior derivation, we thus obtain from (\ref{coexact}),
\begin{equation}
  d J^\pm = d\delta F^\pm,  \;\;\mbox{and}\;\;   * d J^\pm = * d\delta  F^\pm = \pm \,\delta d \,F^\pm   \,.
\end{equation}
It is now a simple matter of combining the two and inverting the de Rahm-Laplace operator $\Delta = d\delta + \delta d $ to express the field strengths in terms of the conserved currents as
\begin{equation}
  \label{FSJ}
  F^\pm = \Delta^{-1} (dJ^\pm \pm *d J^\pm)\,.
\end{equation}
If we want the canonical dimensions of all tensors here to agree with the degree of the corresponding differential form, we identify $ F^+_{\mu\nu} + F^-_{\mu\nu} = m_v \rho_{\mu\nu}  $ in (\ref{FSL}) to write the corresponding actions for the (anti-)selfdual $F^\pm$ of dimension two as
\begin{align}
  S_0^\pm &= \int d^4x  \left\{ \frac{1}{4}  F^\pm_{\mu\nu} F^\pm_{\mu\nu} - \frac{1}{2m_v^2} (\partial_\mu F^\pm_{\mu\nu}) \partial_\sigma F^\pm_{\sigma\nu}  \right\} \nonumber 
\\
  &= \frac{1}{4} \big(F^\pm,F^\pm\big) - \frac{1}{2m_v^2} \big(\delta F^\pm,\delta F^\pm\big) \, ,  \label{SHo}
\end{align}
in the last line also expressed in terms of the Hodge inner product for $p$-forms, e.g.
\begin{equation}
   \big(F,F\big) = \int_M F \wedge *F = \int d^4x \, F_{\mu\nu} F_{\mu\nu}
\end{equation}
for our two-forms $F$ in $\mathds R^4$. The last step now is to likewise rescale our conserved currents, i.e.~the co-exact one-forms $J^\pm = \delta F^\pm$, which are originally of dimension three (they are actually dual to three forms), with a current-field identity, $J^\pm = m_v^2 \, V^\pm$, to express them in terms of one-forms $V^\pm$ of dimension one. In terms of these conserved vector fields we then finally obtain from (\ref{FSJ}),
\begin{align}
  \frac{1}{4} \big(F^\pm,F^\pm\big) &= \frac{m_v^4}{2} \big( V^\pm,\delta \Delta^{-2} d \, V^\pm \big) \, , \label{FF}\\
  \frac{1}{2 m_v^2} \big(\delta F^\pm,\delta F^\pm\big) &= \frac{m_v^2}{2} \big( V^\pm,\delta \Delta^{-1} d \, V^\pm \big) \, . \label{delFdelF}
\end{align}
The corresponding two-point functions for the vector fields $V^\pm$ are transverse as they must, and can now be read off from (\ref{SHo}) with these relations: the mass term from (\ref{FF}) produces in momentum space the transverse mass-like term in (\ref{G2Vects}), i.e.~the second one proportional to $m_v^4/p^2$, and the kinetic term from (\ref{delFdelF}) produces the first term in (\ref{G2Vects}), now actually proportional to $m_v^2$.  

\section{FRG flow equations}\label{sec:flow_equations}

In this appendix we provide further details on the derivation of the FRG flow equations within our present truncations.

The flow equation for the effective potential $U_k(\phi^2)$ is given by
\begin{align}
\partial_k U_k 
&= \frac{k^4}{12 \pi^2}\bigg\{
\frac{1+2\,n_{B}(E_{\sigma,k})}{E_{\sigma,k}}
+\frac{3\,\left(1+2\,n_{B}(E_{\pi,k})\right)}{E_{\pi,k}}\nonumber\\
&-\frac{4\Nf \Nc}{E_{\psi,k}} \big(1-n_{F}(E_{\psi,k}-\mu)-n_{F}(E_{\psi,k}+\mu)\big)\bigg\}\,,
\label{eq:flow_eq_eff_pot}
\end{align}
where $n_{B}$ and $n_{F}$ are the bosonic and the fermionic occupation numbers, and the scale-dependent quasi-particle energies are defined by
\begin{equation}
\label{eq:energies}
E_{\alpha,k}\equiv\sqrt{k^2+m_{\alpha,k}^2}\,, \quad \alpha \in
\{\pi,\sigma,\rho,a_1,\psi\}\,.
\end{equation}
The corresponding mass parameters are given by
\begin{alignat}{2}
\label{eq:masses}
&m_{\pi,k}^2&&=2U_k^\prime(\phi^2)\,,\\
&m_{\sigma,k}^2&&=2U_k^\prime(\phi^2)+4 \phi^2\,U_k^{\prime\prime}(\phi^2)\,,\\
&m_{\rho,k}^2&&=m_{v,k}^2\,,\\
&m_{a_1,k}^2&&=m_{v,k}^2 + g^2 \phi^2\,,\\
&m_{\psi,k}^2&&=h_s^2\phi^2\,,
\end{alignat}
where the derivatives of the scale-dependent effective potential denote derivatives with respect to the chirally invariant $\phi^2=\vec{\pi}^2+\sigma^2$ which are evaluated on the grid in field space when integrating the flow equation for the effective potential. 

Evaluated at the scale-dependent minimum  $\phi_{0,k}^2=\sigma_{0,k}^2$
of $U_k(\phi^2)$ these mass parameters at the same time yield the scale-dependent Euclidean curvature masses as determined from the respective two-point functions $ \Gamma^{(2),E}_{\alpha,k}(p) $ in the limit $p^2\to 0$ which are shown in Fig.~\ref{fig:masses}.

The flow equations for the transverse Euclidean two-point functions of $\rho$ and $a_1$ are given by the projections
\begin{align}
\label{eq:projection_flow_eq_two-point_functions}
\partial_k\Gamma^{(2),T}_k(p) &= \frac{1}{3(N_f^2-1)}\, \Pi^{T}_{\mu\nu}(p) \,
\tr\left(\partial_k \Gamma_{\mu\nu,k}^{(2)}(p)\right) \,,
\end{align}
where these equations are evaluated at fixed value in field space, the minimum at the IR scale $\sigma_{0}$. The structure of the flow of $\Gamma_{\rho/a_1,k}^{(2)}$ is illustrated in \Fig{fig:flow_equations_two-point} and contains regulator derivatives $\partial_k R_{k}$, vertices $\Gamma^{(3)}_k$ and $\Gamma^{(4)}_k$ and regulated Euclidean propagators $D_{k}^E (q)$.

The regulated propagator of particle species $\alpha$ is defined as
\begin{align}
\label{eq:reg_propagator}
D_{\alpha,k}^E (q)\equiv \left(\Gamma^{(2)}_{\alpha,k}(q)+R_{\alpha,k}(q)\right)^{-1}\,,
\end{align}
where we choose three-dimensional analogues of optimized Litim-regulator functions $R_{\alpha,k}(q)$ \cite{Litim:2001up},
\begin{alignat}{3}
&R_{\sigma/\pi,k}&&(q) &&= (k^2-\vec{q\,}^2)\, \Theta (k^2-\vec{\,q}^2 )\,,\label{eqs:regulator_scalar}\\
&R_{\psi,k}&&(q)  &&= \mathrm{i} \slashed{\vec q} \,(\sqrt{k^2/\vec q^{\,2}}-1)
\,\Theta \left(k^2-\vec q^{\,2} \right)\label{eqs:regulator_quarks}\,,\\
&R_{\rho/a_1,k}^{T,L}&&(p) &&=\frac{-m_{0,k}^2}{p^2}\,(k^2-\vec{p\,}^2)\,\Pi^{T,L}_{\mu\nu}(p)\, \Theta (k^2-\vec{\,p}^2 )\label{eqs:regulator_vectors}\, ,
\end{alignat}
mostly as in previous studies. The vector-meson regulator in \Eq{eqs:regulator_vectors} is chosen in a way so that it acts like a scale and momentum-dependent mass term which is added to the vector meson two-point function.

The vertices $\Gamma^{(3)}_k$ and $\Gamma^{(4)}_k$ are extracted from the Ansatz for the effective average action \Eq{eq:lagrangian} and basically contain the couplings $h_v$, $g$ as well as the derivatives of the effective potential. To keep things simple in this qualitative study, the couplings $h_s$, $h_v$ and $g$ are chosen scale-independent as in \cite{Jung:2016yxl}.

We are left with flow equations for the vector-meson masses $m_{v,k}^2$ and $m_{0,k}^2$. The flow of the mass parameter $m_{0,k}^2$ is obtained by projecting the transverse part of the vector-meson two-point function
\begin{align}
\Gamma_{\rho,k}^{(2),T}(p) = \frac{-m_{0,k}^2}{p^2}\,\left(p^2+m_{v,k}^2\right)\,,
\end{align}
where the flow equation for $m_{0,k}^2$ is obtained from 
\begin{align}
\label{eq:flow_m0}
\partial_k &m_{0,k}^2 = -\frac{1}{3(N_f^2-1)} \nonumber\\
&\hspace{-2mm}\times\lim_{p \rightarrow 0} \frac{\partial}{\partial{\vec p}^{\, 2}} \left(p^2\,\Pi_{\mu\nu}^{T}(p)\, \tr\left( \partial_k \Gamma_{\rho,k}^{(2)}(p)\right)_{\mu\nu}\right)\,.
\end{align}
For the flow of $m_{v,k}^2$ we first notice that the transverse flow $ \partial_k\Gamma_{\rho,k}^{(2),T}(p) $ of the vector two-point function is regular for $p\to 0$ and hence 
\begin{align}
\partial_k &\left(m_{v,k}^2 \, m_{0,k}^2\right) = -\frac{1}{3(N_f^2-1)} \nonumber\\
&\times\lim_{p \rightarrow 0} \,
p^2\,\Pi_{\mu\nu}^{T}(p)\, \tr\left( \partial_k \Gamma_{\rho,k}^{(2)}(p)\right)_{\mu\nu} =0\,.
\end{align}
Therefore, one has the simple relation here that
\begin{align}
\partial_k m_{v,k}^2 =-\frac{m_{v,k}^2}{m_{0,k}^2}\, \partial_k m_{0,k}^2\,.
\end{align}
These flow equations for $m_{v,k}^2$ and $m_{0,k}^2$ are usually evaluated at the IR minimum $\sigma_0$, i.e.~when solving the flow equations for the two-point functions in the spirit of our grid-code techniques. The notable exception are the $k$-dependencies of these masses as shown in \Fig{fig:masses} where we have evaluated their flow equations at the $k$-dependent minima $\sigma_{0,k}$ in order to illustrate the scale dependence of the corresponding Euclidean curvature masses.

The flow equations are manipulated and traced using the Mathematica tool FormTracer \cite{Cyrol:2016zqb}.

\section{Analytic continuation and spectral functions}\label{sec:analytic_continuation}

To obtain flow equations for retarded two-point functions from their Euclidean counterparts we use the same analytic continuation procedure on the level of the flow equations that was developed and also used already for example in \cite{Kamikado2014,Tripolt2014,Tripolt2014a,Jung:2016yxl}. 

In a first step we exploit the periodicity of the occupation numbers arising in the equations with respect to imaginary and discrete Matsubara modes $\mathrm{i}p_{0,n}$
\begin{align}
&n_{B}(E\pm\mathrm{i}p_{0,n}) \rightarrow n_{B}(E)\,,\\
&n_F(E\pm\mathrm{i}p_{0,n}) \rightarrow  n_F(E)\,,
\label{eq:frg.continuation1}
\end{align}
and then replace this Euclidean energy $p_0$ with a real continuous frequency $\omega$
\begin{align}
\label{eq:continuation2}
\Gamma^{(2),R}(\omega)=-\lim_{\epsilon\to 0} \Gamma^{(2),E}\left(p_0=-\I(\omega+\I\epsilon)\right)\,.
\end{align}
For the imaginary part of $\Gamma^{(2),R}$ the limit $\epsilon\to 0$ can be taken analytically (described in detail in \cite{Jung:2016yxl}), for the real part of $\Gamma^{(2),R}$ we use $\epsilon=0.1$ or $\epsilon=1$, the $\epsilon$-dependence here is almost negligible. Starting with following initial values at the UV cutoff-scale $\Lambda=1500$ MeV
\begin{alignat}{2}
\label{eq:UV_rho} 
&\Gamma^{(2),R}_{\rho,\Lambda}(\omega)=&& \,m_{0,\Lambda}^2 \,\Big(1-\frac{m_{\rho,\Lambda}^2}{\left(\omega +\I\epsilon\right)^2}\Big)\,,\\
\label{eq:UV_a1} 
&\Gamma^{(2),R}_{a_1,\Lambda}(\omega)=&&\,m_{0,\Lambda}^2 \,\Big(1-\frac{m_{a_1,\Lambda}^2}{\left(\omega+\I\epsilon \right)^2}\Big)\,,
\end{alignat}
the flow equations for $\Gamma^{(2),R}_{\rho,k}$ and $\Gamma^{(2),R}_{a_1,k}$ are solved for the real and imaginary part separately until arriving at the IR scale \mbox{$k=0$ MeV}. The spectral functions are then given by the imaginary parts of the retarded propagators,
\begin{equation}
\rho(\omega)=-\frac{1}{\pi}\,\text{Im}\,D^R(\omega)\,,
\end{equation}
which can be expressed in terms of the real and imaginary parts of the retarded two-point functions,
\begin{equation}
\rho(\omega)=\frac{1}{\pi}\,\frac{\text{Im}\,\Gamma^{(2),R}(\omega)}{\left(\text{Re}\,
	\Gamma^{(2),R}(\omega)\right)^2+\left(\text{Im}\,\Gamma^{(2),R}(\omega)\right)^2}\,.
\end{equation}

\bibliography{QCD}

\end{document}